\documentclass[aps,prl,reprint,longbibliography, superscriptaddresses]{revtex4-2} 
\usepackage{mathrsfs}
\usepackage{textcomp, gensymb}
\usepackage{amsmath,gensymb}
\usepackage{amsfonts}
\usepackage{amssymb}
\usepackage{amsthm}
\usepackage{graphicx}
\usepackage{natbib}
\usepackage{color}
\usepackage{hyperref}
\usepackage{bm}
\usepackage[caption=false]{subfig}
\usepackage{verbatim}
\usepackage{braket}
\usepackage{physics}
\usepackage{nicefrac}

\DeclareMathOperator{\atanh}{arctanh\!}

\newcommand{\nocomments}{\long\def\comm##1\commend{}}

\nocomments

\newcommand{\akcom}[1]{{\comm \color{red} Akash: ``#1'' \commend}}

\newcommand{\ak}[1]{{\color{black} #1}}
\newcommand{\al}[1]{{\color{black} #1}}

\begin{document}
\title{Resolving nonclassical magnon composition of a magnetic ground state via a qubit}
\author{Anna-Luisa E. Römling}
\email{anna-luisa.romling@uam.es}
\affiliation{Condensed Matter Physics Center (IFIMAC) and Departamento de Física
Teórica de la Materia Condensada, Universidad Autónoma de Madrid,
E-28049 Madrid, Spain}

\author{Alejandro Vivas-Viaña}
\affiliation{Condensed Matter Physics Center (IFIMAC) and Departamento de Física
Teórica de la Materia Condensada, Universidad Autónoma de Madrid,
E-28049 Madrid, Spain}
\author{Carlos Sánchez Muñoz}
\affiliation{Condensed Matter Physics Center (IFIMAC) and Departamento de Física
Teórica de la Materia Condensada, Universidad Autónoma de Madrid,
E-28049 Madrid, Spain}
\author{Akashdeep Kamra}
\affiliation{Condensed Matter Physics Center (IFIMAC) and Departamento de Física
Teórica de la Materia Condensada, Universidad Autónoma de Madrid,
E-28049 Madrid, Spain}

\begin{abstract}
	Recently gained insights into equilibrium squeezing and entanglement harbored by magnets point towards exciting opportunities for quantum science and technology, while concrete protocols for exploiting these are needed. Here, we theoretically demonstrate that a direct dispersive coupling between a qubit and a noneigenmode magnon enables detecting the magnonic number states' quantum superposition that forms the ground state of the actual eigenmode - squeezed-magnon - via qubit excitation spectroscopy. Furthermore, this unique coupling is found to enable control over the equilibrium magnon squeezing and a deterministic generation of squeezed even Fock states via the qubit state and its excitation. Our work demonstrates direct dispersive coupling to noneigenmodes, realizable in spin systems, as a general pathway to exploiting the equilibrium squeezing and related quantum properties thereby motivating a search for similar realizations in other platforms.
\end{abstract}
\maketitle

%-----------------------------------------Intro------------------------------------------------------- %
\textit{Introduction.}---Quantum superposition is a central concept and ingredient underlying diverse phenomena from entanglement to the quantum speed up in computing~\cite{wehner2018,laucht2021}. A bosonic mode, such as a photon, can be driven into a so-called nonclassical superposition of its eigenstates - number or Fock states - thereby admitting various quantum advantages~\cite{gerry2005,walls2008}, such as enhancement in its coupling to a qubit via squeezing~\cite{walls1983, leroux2018,qin2018,burd2021,zeytinoglu2017}. At the same time, engineering a dispersive effective interaction $\sim \hat{c}^{\dagger}\hat{c}\hat{\sigma}_{z} $ between the boson (annihilation operator $\hat{c}$) and the qubit $\hat{\sigma}_z$ leads to the latter's excitation frequency becoming multivalued and providing information on the boson's wavefunction~\cite{schuster2007,boissonneault2009,bianchetti2009}. This has been exploited to measure the quantum superposition of the number states that constitutes a given bosonic state~\cite{schuster2007,bianchetti2009,lachance-quirion2017,kono2017,lachance-quirion2019,xu2023}. Since such bosons are also the interconnects in quantum computers, this interplay between their nonclassical states and qubits bears a high relevance for emerging quantum technologies~\cite{laucht2021,terhal2020}.

The bosonic spin excitations of magnets, broadly called magnons, potentially offer advantages in realizing quantum properties~\cite{lachance-quirion2019,kamra2020,yuan2022,awschalom2021}. Magnets have been shown to naturally harbor nonclassical squeezed states in {\em equilibrium}~\footnote{We emphasize the focus of this work on equilibrium squeezed-magnons and their ground state vacuum (e.g., see~\cite{kamra2016,kamra2020,yuan2022,zou2020,wuhrer2022,shim2020,kamra2019,yuan2020,mousolou2021}). These are qualitatively distinct from the squeezed states of magnons generated in nonequilibrium via some drives. There also exists much interest in and excitement about such nonequilibrium quantum states of magnons (e.g., see~\cite{yuan2022,sharma2021,zhao2004,li2019,elyasi2020,kounalakis2022}).} arising from an interplay between energy minimization and the Heisenberg uncertainty principle~\cite{kamra2016,kamra2020,yuan2022}. For example, the ground state and eigenmodes of an anisotropic ferromagnet are constituted by nonclassical superpositions of states with different number of spin flips or, equivalently, magnons~\cite{kamra2020,kamra2017}. The latter are not the eigenmodes but represent the natural or physical basis for the magnet. Hence, the question arises if and how one can measure such nonclassical superpositions of noneigenmode basis states, that constitute the system eigenmodes. An answer to this is also desirable for harnessing the concomitant {\em equilibrium} entanglement harbored by these spin systems for useful quantum information tasks. 

\begin{figure}[tb]
	\centering
	\includegraphics[width=0.7\columnwidth]{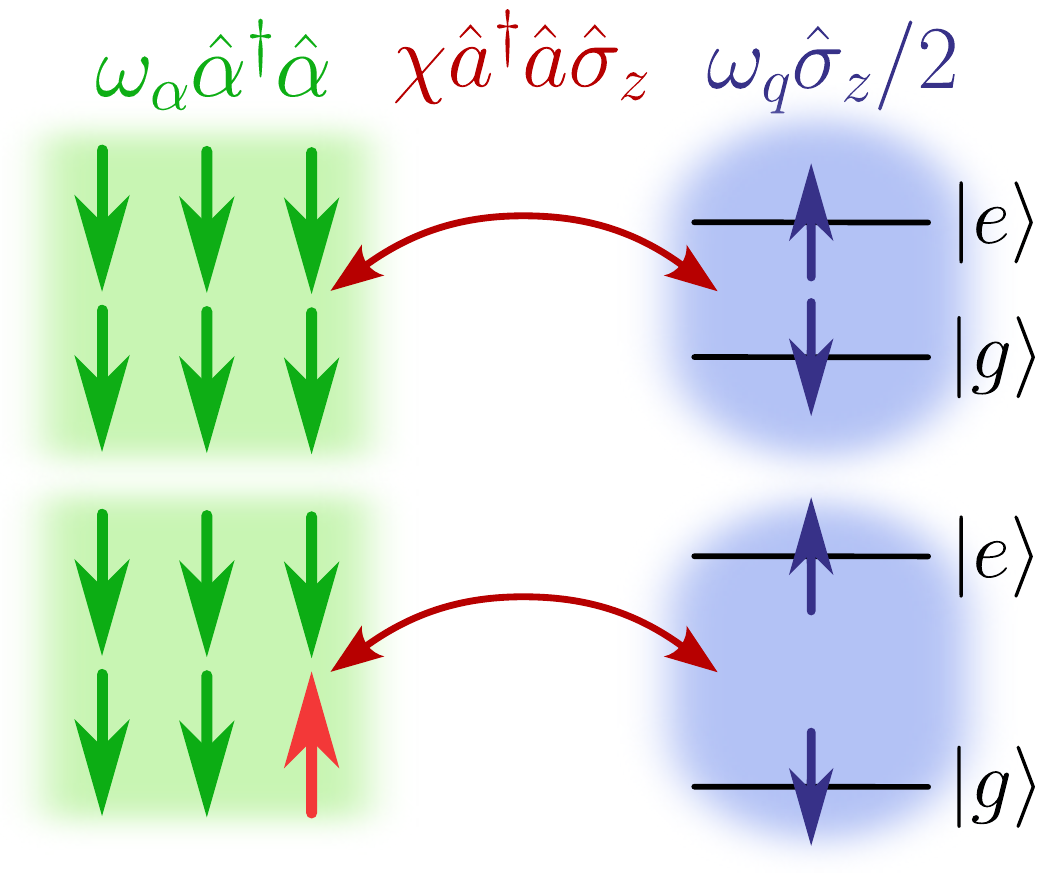}
	\caption{Schematic depiction of the system. The bosonic uniform magnon mode in a ferromagnet (FM, green) is coupled to a spin qubit (blue) through a spin-spin (e.g., exchange) interaction. The ferromagnetic eigenmode is squeezed-magnon $\hat{\alpha}$, while the qubit $\hat{\sigma}_z$ interacts dispersively with the spin-flip or magnon $\hat{a}$ via $\chi \hat{\sigma}_z \hat{a}^{\dagger}\hat{a}$. This direct dispersive coupling originates from the qubit energy depending on the total FM spin, which is governed by the number of spin-flips or magnons (compare upper and lower panels).}\label{fig:Fig1}
\end{figure}

\begin{figure}[tb]
	\centering
	\includegraphics[width=1\columnwidth]{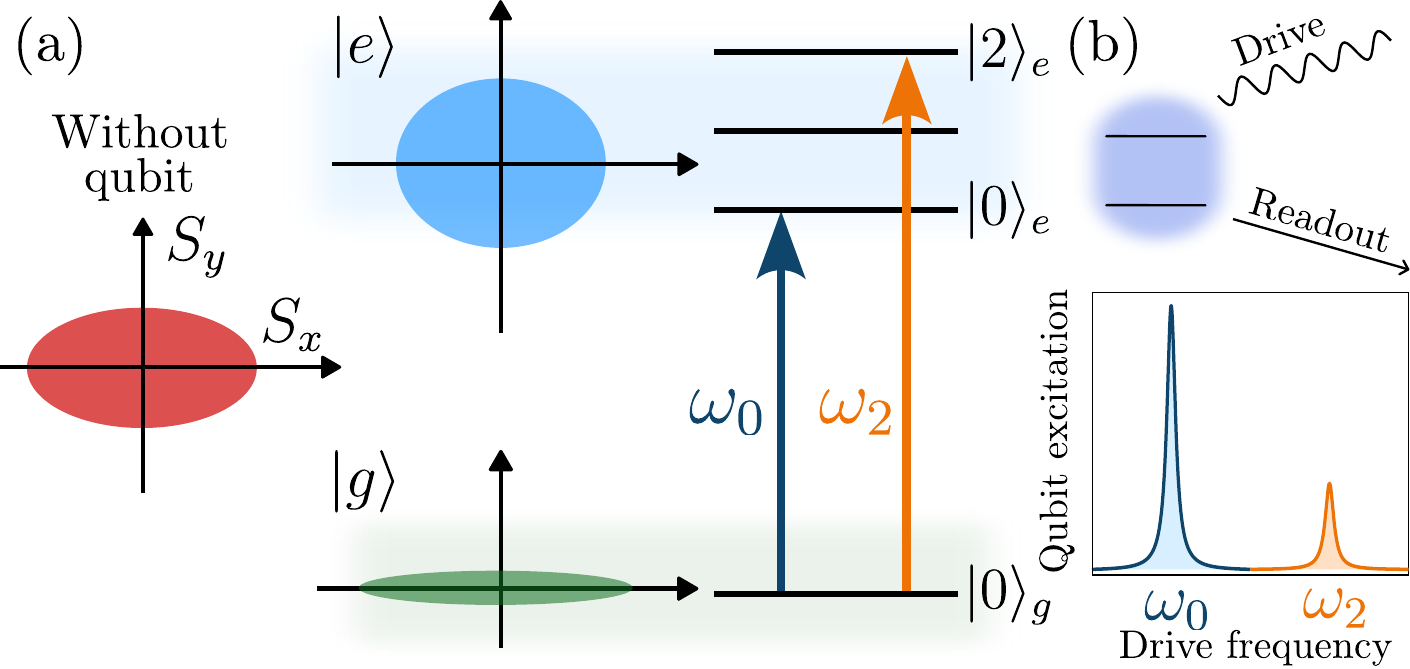} 
	\caption{Qubit excitation spectroscopy of squeezed-magnon vacuum. (a) The ferromagnet (FM) hosts equilibrium-squeezed magnons and corresponding vacuums. As a result, the zero-point quantum fluctuations depicted in the spin phase space bear elliptical profiles~\cite{kamra2020}, indicative of their squeezing. The degree of squeezing is different in three cases: (i) qubit not coupled to the FM (red), (ii) qubit in excited state $\Ket{e}$ (blue), and (iii) qubit in ground state $\Ket{g}$ (green). When one spectroscopically probes the qubit excitation energy $\left( \Ket{g} \to \Ket{e} \right)$, the squeezed-magnon number can change from $0$ to any number state available in the superposition, due to the differing magnon-squeezings in the qubit excited and ground states. (b) This effectively allows to probe the squeezed-magnon vacuum as a superposition of {\em even} magnon number states, with each peak (only first two depicted here) in the qubit excitation spectroscopy measuring a term in the superposition.}
	\label{fig:Fig2}
\end{figure}

In this Letter, taking inspiration from the successful detection of nonequilibrium nonclassical superpositions via a qubit~\cite{schuster2007,lachance-quirion2017,lachance-quirion2019,kono2017,xu2023} and building upon recent advances in probing magnets via qubits~\cite{agarwal2017,chatterjee2019,flebus2018,lachance-quirion2017,lachance-quirion2019,wolski2020,casola2018,xu2023, liu2021}, we address the question posed above. We theoretically demonstrate a protocol for measuring the intrinsic nonclassical superposition that forms the squeezed-magnon vacuum ground state of an anisotropic ferromagnet. We find that the conventional qubit spectroscopy employing a coherent qubit-magnon coupling~\cite{schuster2007,boissonneault2009, tabuchi2015} fails in this goal. However, we show that achieving a direct dispersive interaction (Fig.~\ref{fig:Fig1}) between the qubit and the noneigenmode magnon is the key to achieving this goal. Such a coupling may result from, e.g., the exchange interaction between the magnet and a spin qubit~\cite{burkard2020,chatterjee2021}. Furthermore, our proposed qubit-magnon coupling enables a deterministic protocol to generate nonequilibrium squeezed even Fock states~\cite{nieto1997,kral1990} by driving the qubit at specific frequencies (Fig.~\ref{fig:Fig2}).

%-----------------------------------------Model------------------------------------------------------- %

%Basic set-up/Hamiltonian
\textit{Direct dispersive coupling between magnon and qubit.}---We consider a ferromagnetic insulator with its equilibrium spin order along the z axis and a spatially uniform (wavevector $\bm{k} = \bm{0}$) magnonic mode, represented by the annihilation operator $\hat{a}$. The ferromagnet is coupled to a spin qubit, represented by the operator $\hat{\sigma}_{z}$, via a spin-spin interaction such as dipolar or exchange coupling (Fig.~\ref{fig:Fig1})~\cite{trifunovic2013,skogvoll2021, bender2015, takahashi2010, kamra2016a}. The \ak{$\hat{S}_z\hat{\sigma}_z$} contribution of the spin-spin interaction provides a direct dispersive coupling $\sim \hat{a}^{\dagger}\hat{a}\hat{\sigma}_{z}$ (see Supplemental Material (SM)~\cite{SupplMat}). For the moment, we disregard any coherent coupling returning to it later. Due to magnetic anisotropy in the x-y plane, magnons are not the eigenexcitations~\cite{kamra2016,skogvoll2021} and the total Hamiltonian reads ($\hbar=1$) 
\begin{equation}
	\hat{\mathcal{H}}_{\text{sys}}=A\hat{a}^{\dagger}\hat{a}+B\hat{a}^{2}+B^{*}\hat{a}^{\dagger2}+\frac{\omega_{q}}{2}\hat{\sigma}_{z}+\chi\hat{a}^{\dagger}\hat{a}\hat{\sigma}_{z}\label{eq:H_sys}
\end{equation}
where $A$ and $B$ parametrize the anisotropic ferromagnet~\cite{skogvoll2021} with $B$ resulting from the x-y plane anisotropy, $\omega_{q}$ is the excitation energy of the uncoupled qubit, and $\chi$ (assumed positive here) is the direct dispersive coupling strength. A derivation of Eq.~(\ref{eq:H_sys}) is presented in the SM~\cite{SupplMat}. 

%eigenmodes of the magnet
The ferromagnet only part of the Hamiltonian in Eq.~\eqref{eq:H_sys} can be diagonalized to $\omega_\alpha \hat{\alpha}^\dagger \hat{\alpha}$ with $\hat{\alpha} = \hat{a} \cosh r  + \hat{a}^{\dagger} \sinh r e^{i \theta}$~\cite{kamra2016,skogvoll2021} and 
\begin{align}
	\omega_\alpha &= \sqrt{A^2 - 4\mid B \mid^2}, \label{eq:w_alpha}\\
	2r &= \atanh \left( \frac{2\mid B\mid}{A}\right) \label{eq:r}. 
\end{align}
We refer to the eigenmode $\hat{\alpha}$ as bare squeezed-magnon, since it is related to the magnon $\hat{a}$ via the single-mode squeeze operator~\cite{kamra2016,skogvoll2021,gerry2005}. The squeezing variables $r$ and $\theta$ are determined by $A$ and $B$ of Eq.~\eqref{eq:H_sys} (see SM~\cite{SupplMat} for further details), noting that squeezing and $r$ vanish for $B = 0$. As a result, the ferromagnet ground state is vacuum of the squeezed-magnon $\hat{\alpha}$, which is formed by a quantum superposition of the even magnon $\hat{a}$ number states~\cite{kamra2020,yuan2022}. Since, the $\hat{a}$ magnons are not the eigenmodes, it is not clear how to detect this nonclassical superposition. %As per Eq.~\eqref{eq:H_sys}, the spin qubit couples via a direct dispersive interaction to the magnon $\hat{a}$ (Fig.~\ref{fig:Fig1}), which is not the eigenmode. We now exploit this feature to demonstrate resolution of the internal nonclassical structure of the squeezed-magnon vacuum.

%-----------------------------------------Analytics------------------------------------------------------- %

%Main idea and overview   
\textit{Magnon number dependent qubit excitation energy.}---The nonequilibrium superpositions of eigenmode number states have been investigated via measurement of multiple peaks in a qubit excitation spectroscopy~\cite{schuster2007,bianchetti2009,kono2017}. Here, each peak comes from a different number state contribution to the superposition. Despite a similar motivation, this should be clearly contrasted with our goal and challenge of resolving the noneigenmode magnon number state composition of the equilibrium/eigenmode state - the squeezed-magnon vacuum~\cite{kamra2016,kamra2020,yuan2022}. We hypothesize that the desired resolution can be accomplished in our considered model (Fig.~\ref{fig:Fig1}) when the qubit energy depends directly on the noneigenmode magnon number ($\sim \chi\hat{a}^{\dagger}\hat{a}\hat{\sigma}_{z}$), by spectroscopically probing the qubit excitation energies. We now evaluate the latter to examine this hypothesis.

%Qubit ground and excited states
We first project the total Hamiltonian Eq.~\eqref{eq:H_sys} onto the qubit ground state $\Ket{g}$. The reduced Hamiltonian $\hat{\mathcal{H}}_{g}=\Bra{g}\hat{\mathcal{H}}_{\text{sys}}\Ket{g}$ is obtained as
\begin{align}
	\hat{\mathcal{H}}_{g} & =\left(A-\chi\right)\hat{a}^{\dagger}\hat{a}+B\hat{a}^{2}+B^{*}\hat{a}^{\dagger2}-\frac{\omega_{q}}{2}\,.\label{eq:reduced_Ham_g}
\end{align}
In a direct analogy with the discussion and analysis following Eq.~\eqref{eq:H_sys}, the reduced Hamiltonian Eq.~\eqref{eq:reduced_Ham_g} can be diagonalized to $\omega_\alpha^g \hat{\alpha}_g^\dagger \hat{\alpha}_g$ with a different squeezed-magnon $\hat{\alpha}_{g}$ eigenmode characterized by a frequency $\omega_{\alpha}^{g}<\omega_{\alpha}$ and squeezing factor $r_{g}>r$. $\omega_{\alpha}^{g}$ and $r_{g}$ are obtained from Eqs.~\eqref{eq:w_alpha} and \eqref{eq:r} by substituting $A\rightarrow A-\chi$~\footnote{The ground state stability requires $A \geq 2\left|B\right|+\chi$ which yields a finite $r_{g}$.}. We will refer to $\hat{\alpha}_{g}$ as the ground state squeezed-magnon harboring a different magnetic vacuum as compared to the isolated ferromagnet [Fig.~\ref{fig:Fig2}(a)]. The projection $\hat{\mathcal{H}}_{e}=\Bra{e}\hat{\mathcal{H}}_{\text{sys}}\Ket{e}$
onto the qubit excited state $\Ket{e}$ can be obtained from Eq.~(\ref{eq:reduced_Ham_g})
by changing the sign of $\chi$ and $\omega_q$. Analogous to the discussion above, the bosonic eigenmode of $\hat{\mathcal{H}}_{e}$ becomes the excited state squeezed-magnon $\hat{\alpha}_{e}$ characterized by eigenenergy $\omega_{\alpha}^{e}>\omega_{\alpha}$ and squeezing factor $r_{e}<r$ [Fig.~\ref{fig:Fig2}(a)], with $\omega_{\alpha}^{e}$ and $r_{e}$ obtained from Eqs.~\eqref{eq:w_alpha} and \eqref{eq:r} on replacing $A\rightarrow A+\chi$.

%Summary of eigenstates
Altogether, we have diagonalized our Hamiltonian Eq.~\eqref{eq:H_sys} denoting the eigenstates by $\Ket{n}_{e}$ and $\Ket{n}_g$, where the subscript $g$ or $e$ indicates the qubit state and $n\in\mathbb{N}$ labels the different Fock states. The key point is that the magnonic eigenmodes and their respective squeezing are different in three cases: (i) isolated ferromagnet, (ii) qubit in its ground state, and (iii) qubit in its excited state [see Fig.~\ref{fig:Fig2}(a)]. 

% Transitions on qubit excitation
The typical qubit excitation spectroscopy measures qubit energy corresponding to the transition $\Ket{g} \to \Ket{e}$, while the boson number state remains the same~\cite{boissonneault2009, schuster2007}. Consequently, when we have a nonequilibrium superposition of multiple number states, the result is observation of boson number-dependent qubit energy that manifests itself as multiple spectroscopy peaks. In sharp contrast, our system has a boson mode whose squeezing depends on the qubit state. Hence, the excitation of qubit need not preserve the boson number. Thus, transitions $\Ket{0}_g \to \Ket{n}_e$ will take place with probability $p_n = \left| c_n \right|^2 \equiv \left|{\phantom{|\!\!}}_e\!\braket{n}{0}_g \right|^2$ resulting in correspondingly high spectroscopy peaks. As demonstrated in the SM~\cite{SupplMat}, the ground state $\ket{0}_{g}$ is squeezed with respect to the excited state squeezed-magnon vacuum $\ket{0}_{e}$ with effective squeezing factor of $r_{\text{eff}} = r_{g}-r_{e}$~[Eq.~(\ref{eq:r})].
Thus, we may express $\Ket{0}_{g} = \sum_{n} c_{n}\Ket{n}_{e}$ with~\cite{gerry2005,walls2008}
\begin{align}
	c_{2n} & =\frac{1}{\sqrt{\cosh r_{\text{eff}}}}\left(-e^{i\theta}\tanh r_{\text{eff}}\right)^{n}\frac{\sqrt{\left(2n\right)!}}{2^{n}n!}\label{eq:superposition-probing-basis}
\end{align}
and $c_{2n+1}=0$ for $n\in\mathbb{N}$. To sum up, the qubit spectroscopy should yield a peak for each of the superposition contributions [Fig.~\ref{fig:Fig2}(b)], as intuitively hypothesized above. However, it resolves the ground state squeezed-magnon vacuum $\Ket{0}_g$ in terms of the excited state squeezed-magnon number states $\Ket{n}_e$ [Eq.~\eqref{eq:superposition-probing-basis}].

\begin{figure}[tb]
	\centering
	\includegraphics[width=0.9\columnwidth]{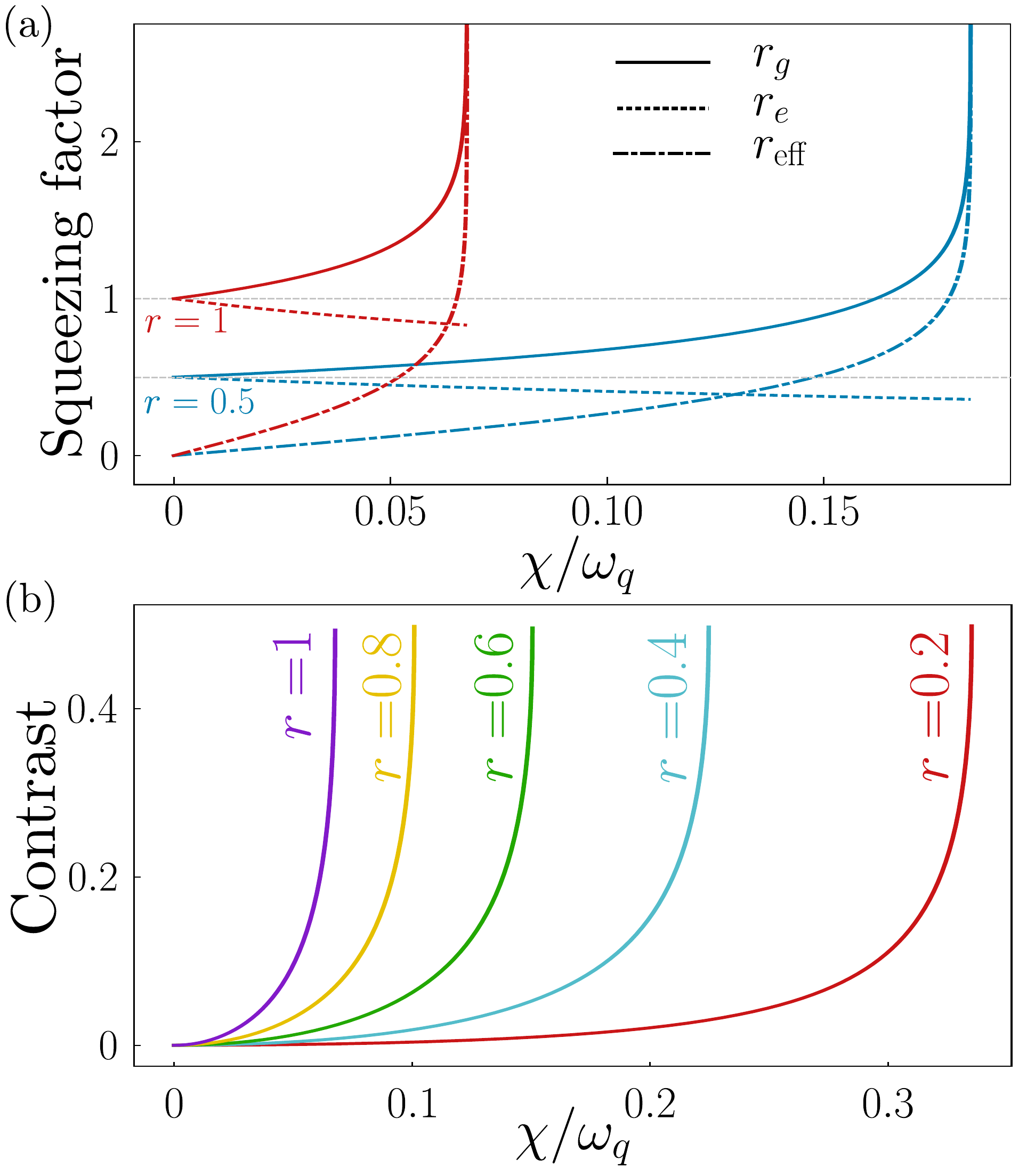} 
	\caption{(a) Squeezing factors {\it vs.} $\chi$ for the magnonic eigenmodes in the qubit ground state
		$r_{g}$ (solid), the qubit excited $r_{e}$ (dotted) and effective squeezing $r_{\text{eff}}=r_{g}-r_{e}$ (dashed) considering bare magnon squeezing of $r = 0.5$ (blue) and $r = 1$ (red). (b) Contrast $c = p_{2}/p_{0}$ [Eq.~\eqref{eq:contrast}] as a function of $\chi$ for several values of the squeezing factor $r$. Its vanishing in the limit $r \to 0$ signifies that more than 1 peak in the spectroscopy is observed only for nonzero magnon squeezing. We consider $\omega_\alpha/\omega_q = 0.5$ here.}
	\label{fig:Fig3}
\end{figure}

In Fig.~\ref{fig:Fig3}(a), we plot the squeezing factors $r_{g}$, $r_{e}$ and $r_{\text{eff}}$
as a function of the dispersive coupling strength $\chi$. Only at a certain value of $\chi$, $r_{\text{eff}}$ is equal to the squeezing $r$ of the bare squeezed-magnon. In this case, the spectroscopy would probe the ``true'' distribution of the bare squeezed-magnon $\hat{\alpha}$ vacuum in terms of the magnon $\hat{a}$ Fock states. Nevertheless, employing our analysis above, a knowledge of $\chi$ and $\omega_{\alpha}$ allows one to translate an observed superposition into any desired basis.

%Excitation energies  
We now examine the position of the spectroscopy peaks. As per energy conservation, the transition $\Ket{0}_g \to \Ket{2n}_e$ occurs when the drive frequency matches the energy difference between the two states. As detailed in the SM~\cite{SupplMat}, this is evaluated as $\omega_{2n}$: 
\begin{equation}
	\omega_{2n}  =\omega_{q}+\frac{\omega_{\alpha}^{e}-\omega_{\alpha}^{g}}{2}-\chi + 2n\cdot\omega_{\alpha}^{e}\,.\label{eq:trans_freq_n}
\end{equation}
For $\chi\ll\min\left[\left|A\right|,\left|\left|B\right|\left(A/2\left|B\right|-2\left|B\right|/A\right)\right|\right]$, Eq.~(\ref{eq:trans_freq_n}) becomes $\omega_{2n}\approx\omega_{q}+2\chi\cdot\sinh^{2}r + 2n\left[\omega_{\alpha}+\chi\cosh\left(2r\right)\right]$.
The different peaks are now well separated by multiples of the bare squeezed-magnon frequency $\omega_{\alpha}$, potentially making them easier to detect~\footnote{In contrast, the conventional qubit spectroscopy~\cite{schuster2007,boissonneault2009} of nonequilibrium superpositions yields peaks separated by the typically smaller quantity $\chi$}. 

%Contrast
In order to guide and quantify the measurability of multiple peaks resulting from the superpositions, we define  ``contrast'' as the ratio $c = p_2/p_0$ evaluating it as
\begin{equation}
	2 c =\tanh[2](r_{\text{eff}}). \label{eq:contrast}
\end{equation}
The contrast $c$, plotted in Fig.~\ref{fig:Fig3}(b), generally characterizes the reduction of subsequent peaks expected in the qubit spectroscopy. For small coupling strengths $\left|\chi\right|\ll\min\left[\left|A\right|,\left|A\left(A/2\left|B\right|-2\left|B\right|/A\right)\right|\right]$, we obtain $c\approx 2\left|B\right|^{2}\chi^{2}/\left(A^{2}-4\left|B\right|^{2}\right)^{2}$. For small $\left|B\right|\ll\min\left[\left|A-\chi\right|,\left|A+\chi\right|\right]$ and thus squeezing, the contrast can be expanded as $c\approx 2\chi^{2}\left|B\right|^{2}/\left(A^{2}-\chi^{2}\right)^{2}$. Thus, the equilibrium superposition peaks can be observed in the qubit spectroscopy when both the direct dispersive interaction strength $\chi$ and squeezing $r$ are nonzero, with the resolvability of the peaks increasing with both these parameters.

%-----------------------------------------Simulations------------------------------------------------------- %

\begin{figure}[tb]
	\centering
	\includegraphics[width=0.9\columnwidth]{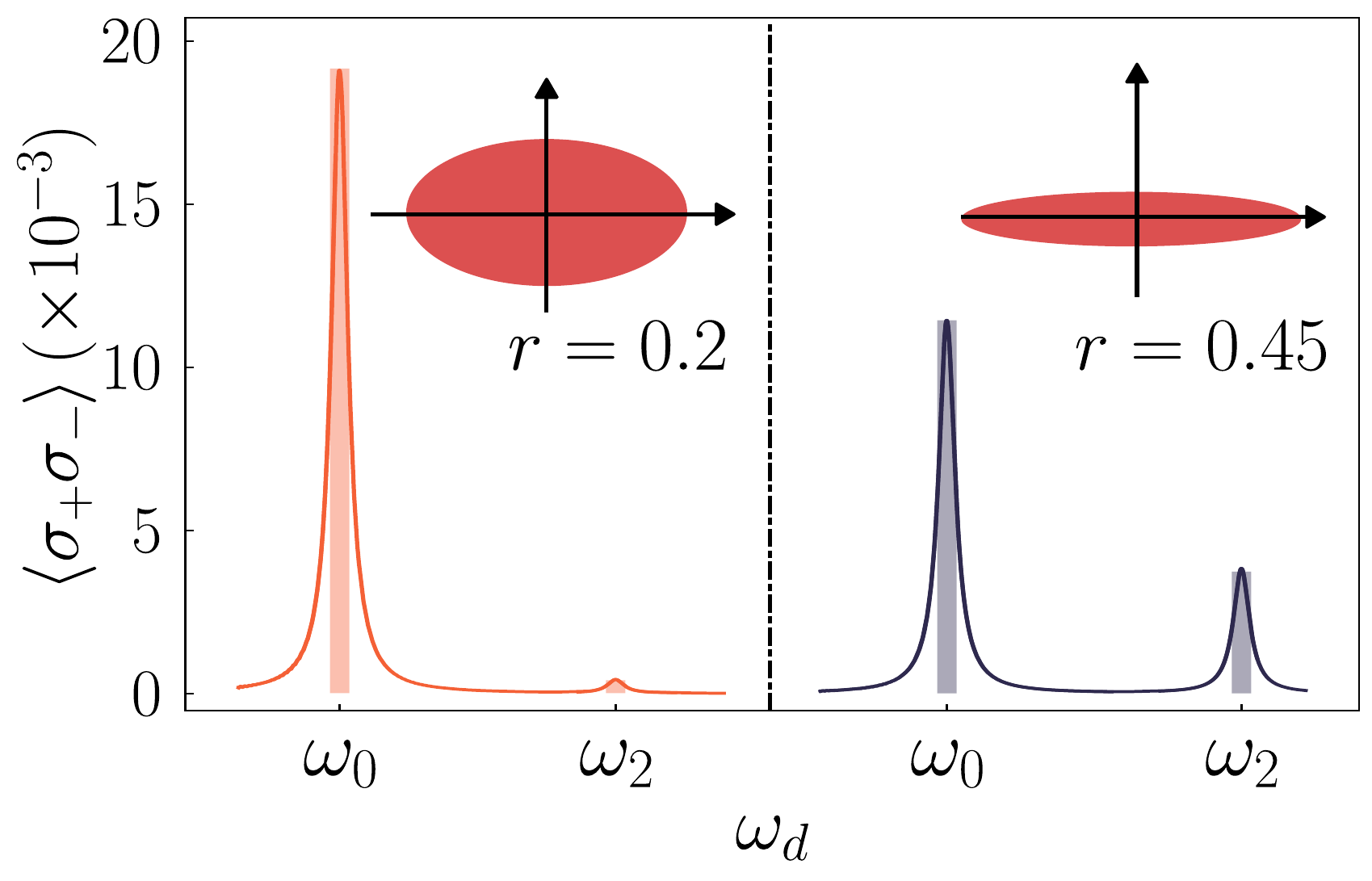} 
	\caption{Numerical simulation of qubit spectroscopy using a Rabi drive. Steady state qubit excitation $\left\langle \hat{\sigma}_{+}\hat{\sigma}_{-}\right\rangle $ is plotted against the Rabi drive frequency $\omega_{d}$ for two different values of bare magnon-squeezing $r$. The first two qubit excitation frequencies $\omega_{0}$ and $\omega_{2}$ are observed. The shaded bars depict the analytically evaluated excitation distributions [Eqs.~\eqref{eq:superposition-probing-basis} and \eqref{eq:trans_freq_n}], underlining their good agreement with the simulations. Parameters employed in the simulation are $\omega_\alpha/\omega_q = 0.5$, $\chi/\omega_q = 0.2$, $\gamma_q/\omega_q = 0.1$ and $\Omega_d/\omega_q = 0.014$. The numerical method is detailed in the SM~\cite{SupplMat}.}
	\label{fig:Fig4}
\end{figure}

%Introduction/Goal%
\textit{Simulation of qubit spectroscopy.}---We now corroborate and complement our analytic considerations above by simulating a qubit spectroscopy setup using the QuTip package~\citep{johansson2012,johansson2013}. While different experimental methods can be employed to probe the qubit excitation energy~\cite{schuster2007, kono2017}, here we consider a microwave qubit drive described by $
\hat{\mathcal{H}}_{\text{d}}=\Omega_{d}\cos\left(\omega_{d}t\right)\left(\hat{\sigma}_{+}+\hat{\sigma}_{-}\right)%, \label{eq:drive}
$
where $\Omega_{d}$ denotes the Rabi frequency quantifying the drive strength, while $\omega_d$ becomes the drive frequency. As detailed in the SM~\cite{SupplMat}, we consider Eq.~\eqref{eq:H_sys} and $\hat{\mathcal{H}}_\text{d}$ to describe our system and account for qubit dissipation~\citep{breuer2007} via one collapse operator $\hat{C}=\sqrt{\gamma_{q}}\hat{\sigma}_{-}$ with qubit decay rate $\gamma_{q}$. Solving the Lindblad master equation~\citep{lindblad1976,gorini1976,breuer2007} numerically, we investigate the steady state qubit excitation  $\left\langle \hat{\sigma}_{+}\hat{\sigma}_{-}\right\rangle$. $\Omega_d$ is chosen small enough \ak{(see SM~\cite{SupplMat} for a quantification of this smallness)} for the qubit excitation to remain small and in the linear regime~\citep{breuer2007,SupplMat}. With this protocol, the qubit excitation should manifest a peak whenever the drive frequency $\omega_d$ is resonant with a qubit excitation transition. 

%Figure Take Aways %
In Fig.~\ref{fig:Fig4}, we show simulations (solid curves) of the qubit spectroscopy for two squeezing factors $r=0.2$ and $r=0.45$, comparing them with our analytic results plotted as bars at $\omega_d = \omega_{2n}$ [Eq. (\ref{eq:trans_freq_n})] with heights $\propto p_{2n} = |c_{2n}|^2$ [Eq.~(\ref{eq:superposition-probing-basis})]. Thus, our analytics agree well with the simulations. We therefore conclude that the first non-trivial peak indeed stems from the equilibrium squeezing. \ak{Due to a large separation ($\sim \omega_{\alpha}$) between the peaks, experiments may further employ higher values of the drive $\Omega_d$ in measuring the smaller peaks. We demonstrate this point explicitly by simulating the $\omega_{4}$ peak in SM~\cite{SupplMat}.} %h_{n}=\left(\Omega_{d}\gamma_{q}\right)^{2}\cdot p_{n}, FWHM = $\gamma_q/2$ %

%-----------------------------------------Coherent coupling------------------------------------------------------- %
\textit{Consideration of coherent coupling.}---Until now, we have considered a magnet coupled to a spin qubit that offers a direct dispersive coupling $\chi$ [Eq.~\eqref{eq:H_sys}], found to be essential for the key phenomena addressed here. We now examine the role of coherent or Rabi interaction~\cite{rabi1937} parameterized by $g$, such that the system Hamiltonian becomes:
\begin{align}
	\hat{\mathcal{H}}_{\text{sys,SC}}=A\hat{a}^{\dagger}\hat{a} & +B\hat{a}^{2}+B^{*}\hat{a}^{\dagger2}+\frac{\omega_{q}}{2}\hat{\sigma}_{z} \nonumber \\
	& +g\left(\hat{a}^{\dagger}+\hat{a}\right)\left(\hat{\sigma}_{+}+\hat{\sigma}_{-}\right). \label{eq:H_Rabi}
\end{align}
This interaction is universally present in qubits, such as with spin~\cite{burkard2020,chatterjee2021} and superconducting qubits~\cite{tabuchi2015,tabuchi2016,liu2021}, while the direct dispersive coupling is not always available. When the boson and qubit are strongly detuned i.e., $g \ll \left|\omega_{q}-\omega_{\alpha}\right|$, the coherent coupling also results in an effective dispersive interaction $\sim \tilde{\chi} \hat{\alpha}^\dagger \hat{\alpha} \hat{\sigma}_z$~\cite{SupplMat,schuster2007,boissonneault2009,zueco2009,gerry2005} which has been exploited in observing nonequilibrium superpositions in terms of the eigenmode number states. It is not clear whether one can employ this effective dispersive coupling to resolve an equilibrium superposition.

Via numerical simulations of qubit spectroscopy employing Eq.~\eqref{eq:H_Rabi} (see SM~\cite{SupplMat}), we find that the effective dispersive interaction $\sim \tilde{\chi} \hat{\alpha}^\dagger \hat{\alpha} \hat{\sigma}_z$ does not resolve the nonclassical magnon composition of the equilibrium squeezed-magnon vacuum. This can be understood a posteriori since such an effective coupling may address only the eigenmodes $\hat{\alpha}$, and not any internal noneigenmodes. Thus, a direct dispersive interaction $\sim \chi \hat{a}^\dagger \hat{a} \hat{\sigma}_z$ offered by, e.g., a spin qubit is needed for resolving equilibrium superpositions. We also show that any influence of the coherent coupling $g$ when employing a spin qubit system can be suppressed via an adequately large detuning $\left|\omega_{q}-\omega_{\alpha}\right|$~\cite{SupplMat,zueco2009}.

%-----------------------------------------Discussion------------------------------------------------------- %

\textit{Discussion.}---In the conventional qubit spectroscopy for dispersively sensing a nonequilibrium quantum superposition of eigenmode Fock states, the peaks are separated in frequency by $\sim \tilde{\chi}$ which is typically small~\cite{schuster2007, bianchetti2009, boissonneault2009}. In our demonstrated protocol for detecting the equilibrium superposition of noneigenmode Fock states, the corresponding peaks are well-separated $\sim \omega_{\alpha}$, which makes it feasible to detect them~\footnote{This can be done by, for example, employing a larger amplitude of the qubit drive since the qubit excitation increases with the drive~\cite{SupplMat}.} even when they are relatively small \ak{(see SM~\cite{SupplMat})}.

\ak{The direct dispersive interaction results from the $\hat{S}_z\hat{\sigma}_z$ term contained in exchange as well as dipolar spin-spin interaction hosted by multiple magnet--spin qubit platforms, as discussed in SM~\cite{SupplMat}. The resulting $\chi$ offered by an exchange-coupled spin qubit can be large $\sim$ GHz} for small size of the magnet (see the SM~\cite{SupplMat}) making our proposal better suited for nanomagnets. Furthermore, detection of the $n$th nontrivial peak in the qubit spectroscopy is accompanied by the transition $\Ket{0}_g \to \Ket{2n}_e$ which provides a new deterministic approach to generate nonequilibrium squeezed Fock states ($\Ket{2n}_{e} = S^{-1}(r_{\mathrm{eff}}) \Ket{2n}_g$~\cite{kamra2016, kamra2020,nieto1997,kral1990}) by driving the qubit.

%-----------------------------------------Conclusion------------------------------------------------------- %
\textit{Conclusion.}---We have theoretically demonstrated how a direct dispersive interaction between a qubit and a noneigenmode boson (here, a magnon) enables detection of the quantum superposition that makes up the actual eigenmodes (here, squeezed-magnon and its vacuum). The same coupling is shown to allow for a control of the equilibrium magnon squeezing and a deterministic generation of squeezed even Fock states via the qubit state and its resonant excitation. Thus, this direct dispersive interaction, readily available in spin systems, opens new avenues for exploiting the equilibrium squeezing and entanglement harbored by magnets. At the same time, our work inspires a search for the realization of direct dispersive interaction in other, such as optical~\cite{ciuti2007} and mechanical, platforms that could enable access to equilibrium superpositions.

\begin{acknowledgments}
	\textit{Acknowledgements.}---We thank Frank Schlawin for valuable discussions. We acknowledge financial support from the Spanish Ministry for Science and Innovation -- AEI Grant CEX2018-000805-M (through the ``Maria de Maeztu'' Programme for Units of Excellence in R\&D) and grant RYC2021-031063-I funded by MCIN/AEI/10.13039/501100011033 and ``European Union Next Generation EU/PRTR''. A. E. R. acknowledges that the project that gave rise to these results received the support of a fellowship from ``la Caixa'' Foundation (ID 100010434). The fellowship code is LCF/BQ/DI22/11940029. C. S. M. acknowledges that the project that gave rise to these results received the support of a fellowship from “la Caixa” Foundation (ID 100010434) and from the European Union’s Horizon 2020 research and innovation programme under the Marie Skłodowska-Curie Grant Agreement No. 847648, with fellowship code LCF/BQ/PI20/11760026, and financial  support from the Proyecto Sinérgico CAM 2020 Y2020/TCS-6545
	(NanoQuCo-CM).
\end{acknowledgments}

\bibliography{Magnon_statistic}

\widetext
\clearpage
%%%%%%%%%% Merge with supplemental materials %%%%%%%%%%
%%%%%%%%%% Prefix a "S" to all equations, figures, tables and reset the counter %%%%%%%%%%
\setcounter{equation}{0}
\setcounter{figure}{0}
\setcounter{table}{0}
\makeatletter
\renewcommand{\theequation}{S\arabic{equation}}
\renewcommand{\thefigure}{S\arabic{figure}}

\begin{center}
	\textbf{\large Supplementary material with the manuscript Resolving nonclassical magnon composition of a magnetic ground state via a qubit by} \\
	\vspace{0.3cm}
	Anna-Luisa E. Römling, Alejandro Vivas-Viaña, Carlos Sánchez Muñoz, and Akashdeep Kamra
	\vspace{0.2cm}
\end{center}

\setcounter{page}{1}

%------------------------------------Hamiltonian-------------------------------$

%\akcom{I have modified the format so that sections and subsections have numbers now. Throughout the text, whenever you refer to a previous section in the SM, now you should refer to the correct section using the adequate number and/or letter.}

\section{Hamiltonian \label{sec:Hamiltonian}}

In this section, we derive the Hamiltonian [Eq.~(1)] describing the dispersively coupled magnon and qubit.
We start by deriving the ferromagnetic Hamiltonian where we also discuss
the squeezing of the isolated magnon. We then focus on the interfacial
exchange-mediated interaction between the magnon and the
spin qubit.

\subsection{Magnon \label{subsec:H_FM}}

For the ferromagnet, we consider the Zeeman energy induced by an external
magnetic field in $z$-direction, the ferromagnetic exchange interaction
between nearest neighbors and a generalized anisotropy term. The total
spin Hamiltonian reads~\cite{skogvoll2021}
\begin{align}
	\hat{\mathcal{H}}_{\text{F}}= %&
	\left|\gamma\right|\mu_{0}H_{0}\sum_{i}\hat{S}_{iz}-J\sum_{\left\langle i,j\right\rangle }\hat{\boldsymbol{S}}_{i}\cdot\hat{\boldsymbol{S}}_{j}%\nonumber \\
	%& 
	+\sum_{i}\left[K_{x}\left(\hat{S}_{ix}\right)^{2}+K_{y}\left(\hat{S}_{iy}\right)^{2}+K_{z}\left(\hat{S}_{iz}\right)^{2}\right], \label{eq:H_F}
\end{align}
with the gyromagnetic ratio $\gamma<0$ , the applied external field
$H_{0}\boldsymbol{e}_{z}$ (with the unit vector in $z$-direction
$\boldsymbol{e}_{z}$ ), the exchange coupling strength $J$. The spin operator at lattice site $i$ is denoted by $\hat{\boldsymbol{S}}_{i}$ and
$\left\langle i,j\right\rangle $ indicates a sum over nearest neighbors.
The magnetic anisotropy is parametrized by the factors $K_{x}$, $K_{y}$
and $K_{z}$~\citep{kamra2016a,skogvoll2021}. In the ground state,
the spins point in the negative $z$-direction. We assume only small
deviations from that state such that we can operate in the spin wave
approximation. This allows us to use the linearized Holstein-Primakoff transformations~\citep{holstein1940} 
\begin{align}
	\hat{S}_{i+} & =\sqrt{2S}\hat{a}_{i}^{\dagger}\label{eq:HP+},\\
	\hat{S}_{iz} & =-S+\hat{a}_{i}^{\dagger}\hat{a}_{i}, \label{eq:HPz}
\end{align}
with $\hat{S}_{i\pm}=\hat{S}_{ix}\pm i\hat{S}_{iy}$ and the spin
magnitude $S$. The bosonic operator $\hat{a}_{i}^{\dagger}$ creates
a local magnon (``spin flip'') at lattice site $i$. The corresponding
Fourier transformation reads 
\begin{align}
	\hat{a}_{i} & =\frac{1}{\sqrt{N_{F}}}\sum_{\boldsymbol{k}}\hat{a}_{\boldsymbol{k}}e^{-i\boldsymbol{k}\cdot\boldsymbol{r}_{i}}, \label{eq:Fourier}
\end{align}
where $N_{F}$ denotes the total number of lattice sites in the ferromagnet.
Using the transformations Eq.~(\ref{eq:HP+}), Eq.~(\ref{eq:HPz})
and Eq.~(\ref{eq:Fourier}), %Skogvoll et al. 
\citet{skogvoll2021}
show that, retaining only the uniform $\boldsymbol{k}=0$ magnon mode
(denoted as $\hat{a}_{\boldsymbol{0}}=\hat{a}$) , the ferromagnetic
Hamiltonian {[}Eq.~(\ref{eq:H_F}){]} transforms into

\begin{align}
	\hat{\mathcal{H}}_{\text{F}} & =A\hat{a}^{\dagger}\hat{a}+B\hat{a}^{2}+B^{*}\hat{a}^{\dagger2},\label{eq:H_mag}
\end{align}
with the constants $A=\left|\gamma\right|\mu_{0}H_{0}+\left(K_{x}+K_{y}-2K_{z}\right)S$
and $B=S\left(K_{x}-K_{y}\right)/2$. This form of the
ferromagnetic Hamiltonian {[}Eq.~(\ref{eq:H_mag}){]} is used in the
main text {[}Eq.~(1){]}.

In the following, we want to discuss the diagonalization of the isolated
ferromagnet $\hat{\mathcal{H}}_{\text{F}}$ {[}Eq.~(\ref{eq:H_mag}){]}
which can be achieved with a Bogoliubov transformation~\citep{holstein1940,bogoljubov1958}
\begin{align}
	\hat{\alpha} & =\hat{a}\cosh r+\hat{a}^{\dagger}e^{i\theta}\sinh r.\label{eq:Bogoliubov}
\end{align}
We denote $\hat{\alpha}$ as the bare squeezed-magnon~\cite{kamra2016}. Defining the one-mode squeeze operator~\citep{walls2008}
\begin{align}
	\hat{S}\left(\xi\right) & =\exp\left[\frac{\xi^{*}}{2}\hat{a}^{2}-\frac{\xi}{2}\left(\hat{a}^{\dagger}\right)^{2}\right], \label{eq:squeeze_op}
\end{align}
the relation between the bare squeezed-magnon $\hat{\alpha}$ and the magnon $\hat{a}$ can be expressed
as 
\begin{align}
	\hat{\alpha}^{\dagger} & =\hat{S}\left(\xi\right)\hat{a}^{\dagger}\hat{S}^{\dagger}\left(\xi\right).
\end{align}
The complex squeezing factor $\xi=r\exp\left(i\theta\right)$ is dictated
by the parameters $A$ and $B$ via 
\begin{align}
	\tanh\left(2r\right) & =\frac{2\left|B\right|}{A},
\end{align}
and the phase 
\begin{align}
	e^{i\theta} & =\frac{B^{*}}{\left|B\right|}.
\end{align}
The diagonalized Hamiltonian becomes
\begin{align}
	\hat{\mathcal{H}}_{\text{F}} & =\omega_{\alpha}\hat{\alpha}^{\dagger}\hat{\alpha}+\frac{\omega_{\alpha}-A}{2}, 
\end{align}
with the resonance frequency 
\begin{align}
	\omega_{\alpha} & =\sqrt{A^{2}-4\left|B\right|^{2}}, 
\end{align}
which requires $\left|A\right|\geq2\left|B\right|$ for stability.

\subsection{Dispersive Interaction \label{subsec:H_dis}}

Here, we consider an interfacial exchange interaction between the spins
of the ferromagnet and the spin qubit. A similar contribution to the Hamiltonian is obtained from dipolar interaction between a magnet and an appropriately located spin qubit~\cite{trifunovic2013,agarwal2017}. The spin Hamiltonian reads~\citep{skogvoll2021}
\begin{align}
	\hat{\mathcal{H}}_{\text{int}} & =J_{\text{int}}\sum_{l}\hat{\boldsymbol{S}}_{l}\cdot\hat{\boldsymbol{s}}_{l}, \label{eq:H_int}
\end{align}
where $l$ label the interfacial site, $J_{\text{int}}$ is the interfacial
coupling strength~\citep{bender2015,kamra2016a,takahashi2010,zhang2012},
$\hat{\boldsymbol{S}}$ denotes the ferromagnetic spin operator and
$\hat{\boldsymbol{s}}$ the spin of the electronic states comprising the
qubit. %Skogvoll et al. 
\citet{skogvoll2021} show that the spin of
the qubit can be written as
\begin{align}
	\hat{\boldsymbol{s}}_{l} & =\frac{\left|\psi_{l}\right|^{2}}{2}\hat{\boldsymbol{\sigma}}, \label{eq:spin-spin-qubit}
\end{align}
where $\psi_{l}$ is the wave function amplitude of the qubit orbital
at position $l$ and $\hat{\boldsymbol{\sigma}} = \hat{\sigma}_x \boldsymbol{e}_x + \hat{\sigma}_y \boldsymbol{e}_y + \hat{\sigma}_z \boldsymbol{e}_z$ denotes the Pauli vector with the Pauli matrices $\hat{\sigma}_{x,y,z}$ and unit vectors $\boldsymbol{e}_{x,y,z}$. Following this, they demonstrate that the interaction
{[}Eq.~(\ref{eq:H_int}){]} can be expressed in the form
\begin{align}
	\hat{\mathcal{H}}_{\text{int}} & =J_{\text{int}}\sum_{l}\left[\hat{S}_{lz}\hat{s}_{lz}+\frac{1}{2}\left(\hat{S}_{l+}\hat{s}_{l-}+\hat{S}_{l-}\hat{s}_{l+}\right)\right], 
\end{align}
and that the term $\propto\hat{S}_{l+}\hat{s}_{l-}+ H.c.$ results
in a coherent magnon-qubit exchange coupling. Focussing on the uniform
$\boldsymbol{k}=\boldsymbol{0}$ mode, they obtain the coherent interaction
\begin{align}
	\hat{\mathcal{H}}_{\text{coh}} & =J_{\text{int}}N_{\text{int}}\left|\psi\right|^{2}\sqrt{\frac{S}{2N_{F}}}\left(\hat{a}^{\dagger}\hat{\sigma}_{-}+\hat{a}\hat{\sigma}_{+}\right), 
\end{align}
with the averaged wavefunction $\left|\psi\right|^{2}=\sum_{l}\left|\psi_{l}\right|^{2}/N_{\text{int}}$.
Note that $N_{\text{int}}$ denotes the number of interfacial sites.
Let's focus on the term $\propto\hat{S}_{lz}\hat{s}_{lz}$ and name
that part of the Hamiltonian $\hat{\mathcal{H}}_{\text{int},zz}$
for now. Using Eq.~(\ref{eq:HPz}), Eq.~(\ref{eq:Fourier}) and the
$z$-component of Eq.~(\ref{eq:spin-spin-qubit}), we find that $\hat{\mathcal{H}}_{\text{int},zz}$
becomes
\begin{align}
	\hat{\mathcal{H}}_{\text{int},zz}= %& 
	-\frac{SJ_{\text{int}}N_{\text{int}}\left|\psi\right|^{2}}{2}\hat{\sigma}_{z}%\nonumber \\
	%&
	+\frac{J_{\text{int}}}{2N_{F}}\sum_{l}\left|\psi_{l}\right|^{2}\sum_{\boldsymbol{k},\boldsymbol{k}^{\prime}}\hat{a}_{\boldsymbol{k}}^{\dagger}\hat{a}_{\boldsymbol{k}^{\prime}}e^{-i\left(\boldsymbol{k}-\boldsymbol{k}^{\prime}\right)\cdot\boldsymbol{r}_{l}}\hat{\sigma}_{z}.
\end{align}
The first term is a renormalization of the qubit frequency and can be
absorbed into the definition of $\omega_{q}$. We focus on the remaining term naming it $\hat{\mathcal{H}}_{\text{dis}}$:
\begin{align}\label{eq:hdis1}
	\hat{\mathcal{H}}_{\text{dis}} & =\frac{J_{\text{int}}}{2N_{F}}\sum_{l}\left|\psi_{l}\right|^{2}\sum_{\boldsymbol{k},\boldsymbol{k}^{\prime}}\hat{a}_{\boldsymbol{k}}^{\dagger}\hat{a}_{\boldsymbol{k}^{\prime}}e^{-i\left(\boldsymbol{k}-\boldsymbol{k}^{\prime}\right)\cdot\boldsymbol{r}_{l}}\hat{\sigma}_{z}. 
\end{align}
A further simplification of this term requires detailed knowledge of the orbital wavefunction $\psi_{l}$ of the spin qubit. This will depend on the physical system being considered. The final result for this term will also differ when one considers dipolar interaction~\cite{trifunovic2013,agarwal2017} instead of the exchange coupling considered herein. In order to obtain an estimate and simplify Eq.~\eqref{eq:hdis1}, we assume $|\psi_l|^2$ to be spatially independent replacing it with its average value $|\psi|^2$. Under this replacement and assuming the ferromagnet to be thin, we may sum over the interfacial sites $l$ obtaining
\begin{align}\label{eq:hdis2}
	\hat{\mathcal{H}}_{\text{dis}} & = \frac{J_{\text{int}} N_{\text{int}} }{2N_{F}}  \left|\psi\right|^{2} \sum_{\boldsymbol{k},\boldsymbol{k}^{\prime}}\hat{a}_{\boldsymbol{k}}^{\dagger}\hat{a}_{\boldsymbol{k}^{\prime}}\delta_{\bm{k},\bm{k}^\prime} \hat{\sigma}_{z} = \frac{J_{\text{int}} N_{\text{int}} }{2N_{F}}  \left|\psi\right|^{2} \sum_{\boldsymbol{k}}\hat{a}_{\boldsymbol{k}}^{\dagger}\hat{a}_{\boldsymbol{k}} \hat{\sigma}_{z}, 
\end{align}
where $\delta_{\bm{k},\bm{k}^\prime}$ is the Kronecker delta function. Focusing on only the uniform $\bm{k} = \bm{0}$ mode, we obtain the direct dispersive interaction considered in the main text: 
\begin{align}\label{eq:hdisfinal}
	\hat{\mathcal{H}}_{\text{dis}} & =\frac{J_{\text{int}}N_{\text{int}}\left|\psi\right|^{2}}{2N_{F}} ~ \hat{a}^{\dagger}\hat{a}\hat{\sigma}_{z} \equiv \chi ~ \hat{a}^{\dagger}\hat{a}\hat{\sigma}_{z}, 
\end{align}
where we again use the notation $\hat{a}$ for representing the uniform mode. 

\section{Excitation Energies \label{sec:exc_energies}}

As discussed in the main text, \ak{the transition of qubit from its ground to excited state can correspond to} multiple excitation frequencies $\omega_{2n}$ [Eq.~(4)] when coupled to an anisotropic ferromagnet.
In this section, we present the mathematical details on how to obtain
the ground and excited states as well as the transition frequencies
between them.

Let's project the full system Hamiltonian $\hat{\mathcal{H}}_\text{sys}$ {[}Eq.~(1){]}
onto the qubit ground state $\Ket{g}$ and excited state $\Ket{e}$. We
denote the reduced Hamiltonians by $\hat{\mathcal{H}}_{m}^{g}=\Bra{g}\hat{\mathcal{H}}_{\text{sys}}\Ket{g}$
and $\hat{\mathcal{H}}_{m}^{e}=\Bra{e}\hat{\mathcal{H}}_{\text{sys}}\Ket{e}$
respectively and obtain
\begin{align}
	\hat{\mathcal{H}}_{g} & =\left(A-\chi\right)\hat{a}^{\dagger}\hat{a}+B\hat{a}^{2}+B^{*}\hat{a}^{\dagger2}-\frac{\omega_{q}}{2}\label{eq:H_g}, \\
	\hat{\mathcal{H}}_{e} & =\left(A+\chi\right)\hat{a}^{\dagger}\hat{a}+B\hat{a}^{2}+B^{*}\hat{a}^{\dagger2}+\frac{\omega_{q}}{2}.\label{eq:H_e}
\end{align}
%Both operators have the form of a squeezed magnon Hamiltonian. 
We diagonalize Eq.~(\ref{eq:H_g}) and Eq.~(\ref{eq:H_e}) with the help
of Bogoliubov transformations (that have the same form as Eq.~(\ref{eq:Bogoliubov})) and obtain
\begin{align}
	\hat{\mathcal{H}}_{g} & =\omega_{\alpha}^{g}\hat{\alpha}_{g}^{\dagger}\hat{\alpha}_{g}-\frac{\omega_{q}}{2}+\frac{\omega_{\alpha}^{g}-\left(A-\chi\right)}{2}, \\
	\hat{\mathcal{H}}_{e} & =\omega_{\alpha}^{e}\hat{\alpha}_{e}^{\dagger}\hat{\alpha}_{e}+\frac{\omega_{q}}{2}+\frac{\omega_{\alpha}^{e}-\left(A+\chi\right)}{2}.
\end{align}
We denote the eigenmode of Eq.~(\ref{eq:H_g}) as the ground state squeezed magnon $\hat{\alpha}_g$ and the eigenmode of Eq.~(\ref{eq:H_e}) as the excited state squeezed-magnon $\hat{\alpha}_e$. They can be obtained by applying the one-mode squeeze operator $\hat{S}$ [Eq.~\ref{eq:squeeze_op}]
\begin{align}
	\hat{\alpha}_{g/e}^{\left(\dagger\right)} & =\hat{S}\left(\xi_{g/e}\right)\hat{a}^{\left(\dagger\right)}\hat{S}^{\dagger}\left(\xi_{g/e}\right),
\end{align}
with the complex squeezing factors $\xi_{g/e}=r_{g/e}\exp\left(i\theta\right)$.
The absolute value of the squeezing factors read
\begin{align}
	r_{g/e} & =\frac{\text{arctanh}\left(\frac{2\left|B\right|}{A\mp\chi}\right)}{2}, \label{eq:g/e_squeezing}
\end{align}
such that $r_{g}>r$ and $r_{e}<r$, considering $\chi > 0$ as assumed in this work. The angle $\theta$ is given
by $\exp\left(i\theta\right)=B^{*}/\left|B\right|$ and
is therefore the same in both cases. The eigenenergies $\omega_{\alpha}^{g}$
and $\omega_{\alpha}^{e}$ read
\begin{align}
	\omega_{\alpha}^{g/e} & =\sqrt{\left(A\mp\chi\right)^{2}-4\left|B\right|^{2}}, \label{eq:gs_resonance}
\end{align}
such that $\omega_{\alpha}^{g}<\omega_{\alpha}$ and $\omega_{\alpha}^{e}>\omega_{\alpha}$.
Note that the ground state squeezed-magnon requires $A>\chi$ and $A\geq2\left|B\right|+\chi$
for stability. 

The ground state is given by the lowest energy state
of $\hat{\mathcal{H}}_{g}$ which is the ground state squeezed-magnon vacuum $\ket{0}_{g}$.
The ground state energy is therefore 
\begin{align}
	\omega_{0}^{g} & =-\frac{\omega_{q}}{2}+\frac{\omega_{\alpha}^{g}-\left(A-\chi\right)}{2}.
\end{align}
As discussed in the main text, the qubit can be excited into the excited state squeezed-magnon Fock states $\ket{2n}_{e}$ (eigenstates of $\hat{\mathcal{H}}_{e}$). The energy
of an excited state $\ket{2n}_{e}$ reads 
\begin{align}
	\omega_{2n}^{e} & =2n\cdot\omega_{\alpha}^{e}+\frac{\omega_{q}}{2}+\frac{\omega_{\alpha}^{e}-\left(A+\chi\right)}{2}, 
\end{align}
where $2n= {\phantom{|\!\!}}_e\!\bra{2n}\hat{\alpha}_{e}^{\dagger}\hat{\alpha}_{e}\ket{2n}_{e}$.
The excitation energy from the ground state $\ket{0}_{g}$ to an excited
state $\Ket{2n}_{e}$ is then given by the energy difference $\omega_{2n}=\omega_{2n}^{e}-\omega_{0}^{g}$
which explicitly reads
\begin{align}
	\omega_{2n} & =\omega_{q}+\frac{\omega_{\alpha}^{e}-\omega_{\alpha}^{g}}{2}-\chi+2n\cdot\omega_{\alpha}^{e}.
\end{align}
This is Eq.~(4) from the main text.

\section{Effective Squeezing \label{sec:BCH}}

As discussed in the main text, \ak{the qubit coupled to the magnon mode can transition from its ground state $\Ket{0}_{g}$ into different excited states $\Ket{2n}_{e}$ corresponding to the different occupancy $n$ of the magnon mode in the final state}. This is because the squeezing of the ground state squeezed-magnon $r_{g}$ and the excited state squeezed-magnon $r_{e}$ are different. We state in the main text that the ground state squeezed-magnon vacuum $\ket{0}_g$ and the excited state squeezed-magnon vacuum $\ket{0}_e$ are squeezed with respect to each other by an effective squeezing factor $r_{\text{eff}}=r_{e}-r_{g}$. This allows us to write the ground state squeezed-magnon vacuum
$\Ket{0}_{g}$ as a superposition of excited state squeezed-magnon Fock states $\Ket{2n}_{e}$.
In this section, we want to derive the effective squeezing in mathematical
detail. 

It is necessary to express the ground state $\ket{0}_g$ and the excited states $\ket{2n}_e$
in a common basis. We denote the magnon Fock states by $\ket{n}$. This way,
the ground state squeezed-magnon vacuum can be written as
\begin{align}
	\Ket{0}_{g} & =\hat{S}\left(\xi_{g}\right)\Ket{0}\label{eq:gs_magnon_fock}
\end{align}
and the excited state squeezed-magnon Fock states~\citep{kamra2020}
\begin{align}
	\Ket{2n}_{e} & =\hat{S}\left(\xi_{e}\right)\Ket{2n}.
\end{align}
The squeeze operators $\hat{S}\left(\xi_{g/e}\right)$ have the
same form as Eq.~(\ref{eq:squeeze_op}) with the squeezing factors
$\xi_{g/e}$ {[}Eq.~(\ref{eq:g/e_squeezing}){]}. In order to eliminate
the magnon vacuum $\ket{0}$ from Eq.~(\ref{eq:gs_magnon_fock}) and establish
a connection between the squeezed-magnon vacuum states $\Ket{0}_{g}$ and $\Ket{0}_{e}$, we use the relation 
\begin{align}
	\Ket{0} & =\hat{S}\left(-\xi_{e}\right)\Ket{0}_{e}.\label{eq:magnon-vac}
\end{align}
Note that we exploited the fact that the inverse of the squeezing
operator {[}Eq.~(\ref{eq:squeeze_op}){]} is $\hat{S}^{-1}\left(\xi_{e}\right)=\hat{S}\left(-\xi_{e}\right)$.
Inserting Eq.~(\ref{eq:magnon-vac}) into Eq.~(\ref{eq:gs_magnon_fock}),
we find a relationship between the squeezed-magnon vacuum states $\Ket{0}_{g}$
and $\Ket{0}_{e}$ via
\begin{align}
	\Ket{0}_{g} & =\hat{S}\left(\xi_{g}\right)\hat{S}\left(-\xi_{e}\right)\Ket{0}_{e}.
\end{align}
Since the operator $\hat{S}\left(\xi_{g}\right)\hat{S}\left(-\xi_{e}\right)$
is a product of two matrix exponentials {[}see Eq.~(\ref{eq:squeeze_op}){]},
we proceed by evaluating the corresponding Baker-Campbell-Hausdorff
formula~\citep{achilles2012} 
\begin{align}
	e^{\hat{X}}e^{\hat{Y}} & =e^{\hat{Z}}, 
\end{align}
with the operators
\begin{align}
	\hat{X} & =\frac{\xi_{g}^{*}}{2}\hat{a}^{2}-\frac{\xi_{g}}{2}\left(\hat{a}^{\dagger}\right)^{2}, \\
	\hat{Y} & =\frac{\xi_{e}}{2}\left(\hat{a}^{\dagger}\right)^{2}-\frac{\xi_{e}^{*}}{2}\hat{a}^{2}, 
\end{align}
and $\hat{Z}=\hat{X}+\hat{Y}+\frac{1}{2}\left[\hat{X},\hat{Y}\right]+\ldots$.
In order to determine the exponential $\hat{Z}$, we need to evaluate
the commutator $\left[\hat{X},\hat{Y}\right]$ and obtain
\begin{align}
	\left[\hat{X},\hat{Y}\right] & =\frac{\xi_{g}^{*}\xi_{e}-\xi_{g}\xi_{e}^{*}}{4}\left[\hat{a}^{2},\left(\hat{a}^{\dagger}\right)^{2}\right].
\end{align}
The factor in front of the bosonic commutator can be written in polar
representation as
\begin{align}
	\xi_{g}^{*}\xi_{e}-\xi_{g}\xi_{e}^{*} & =r_{g}r_{e}\left(e^{-i\left(\theta_{g}-\theta_{e}\right)}-e^{i\left(\theta_{g}-\theta_{e}\right)}\right),
\end{align}
which becomes zero if $\theta_{g}=\theta_{e}$. This is true in our
case, since the angle $\theta$ is only defined by $B$ which is the
same for both squeezed-magnons, hence $\theta_{g}=\theta_{e}$. Therefore,
the exponential $\hat{Z}$ is simply $\hat{Z}=\hat{X}+\hat{Y}$ and
the product of squeeze operators becomes 
\begin{align}
	\hat{S}\left(\xi_{g}\right)\hat{S}\left(-\xi_{e}\right) & =\exp\left[\frac{\left(\xi_{g}-\xi_{e}\right)^{*}}{2}\hat{a}^{2}-\frac{\xi_{g}-\xi_{e}}{2}\left(\hat{a}^{\dagger}\right)^{2}\right],
\end{align}
which is equal to the squeeze operator $\hat{S}\left(\xi_{g}-\xi_{e}\right)$.
Let us denote this as an effective squeezing $\xi_{\text{eff}}=\left(r_{g}-r_{e}\right)\exp\left(i\theta\right)$
and express the ground state squeezed-magnon vacuum as
\begin{align}
	\Ket{0}_{g} & =\hat{S}\left(\xi_{\text{eff}}\right)\Ket{0}_{e}.
\end{align}
This is the relation we use to obtain Eq.~(3) from the main text.

\section{Qubit spectroscopy simulation \label{sec:simulation_supp}}

In this section, the technicalities of the qubit spectroscopy
are described in more detail. For the spectroscopy, we aim to simulate
the dynamics of the dispersively coupled qubit and magnon. For this,
we consider Lindblad Master equations~\citep{lindblad1976, breuer2007} and implement
a coherent microwave drive for the qubit, such that the full Hamiltonian
used for the simulation is time dependent and reads 
\begin{align}
	\hat{\mathcal{H}}_{\text{sim}}=
	A\hat{a}^{\dagger}\hat{a}+B\hat{a}^{2}+B^{*}\hat{a}^{\dagger2}+\frac{\omega_{q}}{2}\hat{\sigma}_{z}+\chi\hat{a}^{\dagger}\hat{a}\hat{\sigma}_{z} 
	+\Omega_{d}\cos\left(\omega_{d}t\right)\left(\hat{\sigma}_{+}+\hat{\sigma}_{-}\right).
\end{align}
The qubit is considered to be an open quantum system with the Lindblad
dissipator $\hat{C}=\sqrt{\gamma_{q}}\hat{\sigma}_{z}$ and decay
rate $\gamma_{q}$. In order to perform the numerics, the Python
package QuTip~\citep{johansson2012,johansson2013} is used. For the
measurement simulations, we initialize the system in the ground state
of the system Hamiltonian $\hat{\mathcal{H}}_{\text{sys}}$ {[}Eq.
(1){]} and drive the qubit with the frequency $\omega_{d}$. Then,
we let the system evolve until the time $T=15/\gamma_{q}$
and measure the qubit response $\left\langle \hat{\sigma}_{+}\hat{\sigma}_{-}\right\rangle $.
Note that the time evolution is calculated with numerical integration
and the time $T$ is chosen large enough such that the system reaches
a steady state. To account for possible oscillations, the qubit excitation
$\left\langle \hat{\sigma}_{+}\hat{\sigma}_{-}\right\rangle $ is
averaged in the time interval $t\in\left[t_{0},T\right]$ with $t_{0}=14/\gamma_{q}$.
In that interval, we take $N_{\text{step}}$ integration steps such
that the discrete time steps fulfill $t_{n+1}=t_{n}+\Delta t$ with
$\Delta t=\left(N_{\text{step}}\gamma_{q}\right)^{-1}.$ The numerical
value of the steady state qubit excitation can now be expressed as
\begin{align}
	\left\langle \hat{\sigma}_{+}\hat{\sigma}_{-}\right\rangle _{\text{st}} & =\frac{1}{N_{\text{step}}}\sum_{t_{n}}^{T}\left\langle \hat{\sigma}_{+}\hat{\sigma}_{-}\right\rangle \left(t_{n}\right).
\end{align}
The qubit drive $\omega_{d}$ is swept over a frequency range (in
the main text, we choose the range such that the frequencies $\omega_{0}$
to $\omega_{2}$ {[}Eq.~(4){]} are covered). For each drive frequency
$\omega_{d}$, we plot the steady state qubit excitation $\left\langle \hat{\sigma}_{+}\hat{\sigma}_{-}\right\rangle _{\text{st}}$
resulting in the desired qubit spectroscopy with excitation peaks.
This allows us to probe the qubit excitation frequencies. 

Whenever we perform a simulation with the direct dispersive coupling
$\chi$, we fix the simulation parameters of the bare squeezed-magnon
frequency $\omega_{\alpha}/2\pi=5\,\mathrm{GHz}$, the
qubit frequency $\omega_{q}/2\pi=10\,\mathrm{GHz}$, the
qubit dissipation $\gamma_{q}/2\pi=1\,\mathrm{GHz}$ and
the Rabi frequency $\Omega_{d}/2\pi=0.14\,\mathrm{GHz}.$
This way, we have a detuning between the bare squeezed-magnon and qubit of $\Delta=5\,\mathrm{GHz}$.
We choose the simulation parameters in a way that does not disturb
the magnon state and allows for a perturbative treatment. For instance,
the Rabi frequency $\Omega_{d}$ is small enough such that we operate
in the linear regime and do not drive the magnon through the qubit.
Furthermore, we choose the combination of parameters ($\omega_{\alpha},$$\gamma_{q}$,
$\Omega_{d}$), such that peaks are well separated and resolvable. The simulation in Fig.~4 are generated with a direct disperive coupling strength of $\chi = 2\,\mathrm{GHz}$.

\al{
	
	\ak{\section{Resolving Excitation Peaks \label{sec:resolving_peaks}}}
	
	\akcom{Please read and correct typos from your text before sending it out for comments. There were too many of them.}
	
	In this section, we want to discuss \ak{some} limitations of the qubit spectroscopy measurement protocol. We first explore a wider range of Rabi frequency $\Omega_d$. This way, we test if the system \ak{is driven} out of equilibrium and therefore \ak{determine the maximum values of} $\Omega_d$ that allow to probe the magnetic ground state. We then \ak{proceed} to discuss the experimental subtleties when resolving further peaks around resonance frequencies $\omega_4$, $\omega_6$~[Eq.~(6)] and so on. Since the corresponding transitions into the states $\ket{2n}_e$ are suppressed by a factor of $p_{2n}=\left| c_{2n} \right|^2$~[Eq.~(5)] we \ak{outline experimental strategies that may enable resolving these weak and well-separated peaks}. 
	
	\subsection{Linear response and weak Rabi drive \label{subsec:rabi_drive}}
	
	As explained \ak{in section \ref{sec:simulation_supp} above}, we choose the Rabi frequency $\Omega_d$ small enough such that the measurement process is only a perturbation on the system state. By driving the qubit at sufficiently large Rabi frequencies $\Omega_d$, the magnon is driven through the qubit reaching a non-equilibrium state. \akcom{Please avoid too much repetition.}
	
	\ak{An approximate} analytical expression for each of the resonance peaks can be achieved by viewing each transition $\ket{0}_g \rightarrow \ket{2n}_e$ as a two level system, labeled by lowering operator $\hat{\sigma}_{2n}$, and calculating the steady-state qubit population. The master equation \cite{breuer2007} that we employ for this purpose is \akcom{I am leaving it on you to take full responsibility for this analytics since I do not have the time to look into this.}
	\begin{equation}
		\partial_t \hat{\rho}_{2n} = - i\left[ \left( \omega_{2n} - \omega_d  \right)\hat{\sigma}_{2n}^\dagger \hat{\sigma}_{2n} + \frac{\Omega_d}{2} \left( c_{2n}^* \hat{\sigma}_{2n} + c_{2n} \hat{\sigma}_{2n}^\dagger \right), \hat{\rho}_{2n}\right] + \gamma_q \mathcal{D} \left[\hat{\sigma}_{2n}\right] \left( \hat{\rho}_{2n} \right),	
	\end{equation}
	where \ak{$\hat{\rho}_{2n}$} is the corresponding density matrix and $\mathcal{D} \left[\hat{\sigma}_{2n}\right] \left( \hat{\rho}_{2n} \right)$ denotes the dissipator $\mathcal{D} \left[\hat{\sigma}_{2n}\right]\left(\hat{\rho}_{2n}\right) = \hat{\sigma}_{2n} \hat{\rho}_{2n} \hat{\sigma}_{2n}^{\dagger} - \frac{1}{2} \left\{ \hat{\sigma}_{2n}^{\dagger} \hat{\sigma}_{2n}, \hat{\rho}_{2n}\right\}$. The factors $c_{2n}$ [Eq.~(5)] take into account that the transition $\ket{0}_g \rightarrow \ket{2n}_e$ is suppressed by $p_{2n} = |c_{2n}|^2$. The steady state solution of the qubit population is then given by~\cite{walls2008} \akcom{Please cite a reference here, if this expression has been obtained in some other text.}
	\begin{equation}\label{eq:qustead}
		\langle \hat{\sigma}_{2n}^\dagger\hat{\sigma}_{2n} \rangle_{\mathrm{st}} = p_{2n}\frac{\left(\Omega_d/2\right)^2 }{\left(\gamma_q/2 \right)^2 + \left(\omega_{2n} - \omega_d \right)^2 + 2p_{2n} \left(\Omega_d/2\right)^2 }.
	\end{equation}
	\ak{When} $ \Omega_d^2 \ll \gamma_q^2/2 p_{2n}$ is fulfilled, the qubit response scales with $\Omega_d^2$. This is what we denote as the linear response regime. 
	
	In Fig.~\ref{fig:Fig_supp1}, we show the contrast resulting from simulations for a wide range of Rabi frequencies $\Omega_d \in \gamma_q \cdot [10^{-2}, 10^0]$ for three different values of direct dispersive coupling strength  $\chi/\omega_q = 0.15$, $\chi/\omega_q = 0.2$ and $\chi/\omega_q = 0.25$. For comparison, we also plot the theory value expected from Eq.~(7). For small $\Omega_d$ the contrast from simulations is constant with increasing $\Omega_d$ and corresponds to the theory value. For large $\Omega_d > \gamma_q/10$ \ak{corresponding to the considered value of $r$}, the contrast increases and is therefore not representative of the magnetic ground state composition marking the limit of the linear response regime \ak{quantified by Eq.~\eqref{eq:qustead} above}.

	In Fig.~\ref{fig:Fig_supp2} we plot \ak{the average ground state squeezed-magnon number $\langle \hat{\alpha}_g^\dagger \hat{\alpha}_g \rangle_{\mathrm{st}} $ as well as }the steady-state qubit population $\langle \sigma_+ \sigma_- \rangle_{\mathrm{st}}$ at resonance \ak{frequencies} $\omega_0$ and $\omega_2$. The average ground state squeezed-magnon number $\langle \hat{\alpha}_g^\dagger \hat{\alpha}_g \rangle_{\mathrm{st}}$ is almost zero at small $\Omega_d$, \ak{in \al{consistence} with the magnon system being in its ground state}. Due \ak{to} the definition of $\hat{\alpha}_g$ and the excitation into the $\ket{0}_e$ and $\ket{2}_e$\ak{, the} expectation values ${\phantom{|\!\!}}_e\!\bra{0} \hat{\alpha}_g^\dagger \hat{\alpha}_g \ket{0}_e =  \sinh[2](r_\mathrm{eff})$ and  ${\phantom{|\!\!}}_e\!\bra{2} \hat{\alpha}_g^\dagger \hat{\alpha}_g \ket{2}_e = 2\cosh[2](r_\mathrm{eff}) +  3\sinh[2](r_\mathrm{eff})$ are non-zero giving rise to the non-equilibrium steady-state when \ak{the Rabi drive becomes large. Consistent with our conclusion above, maintaining a weak Rabi drive in the linear regime [Eq.~\eqref{eq:qustead}] ensures that the magnon system maintains its ground state.}
	
	\begin{figure}[tb]
		\centering\includegraphics[width=0.6\columnwidth]{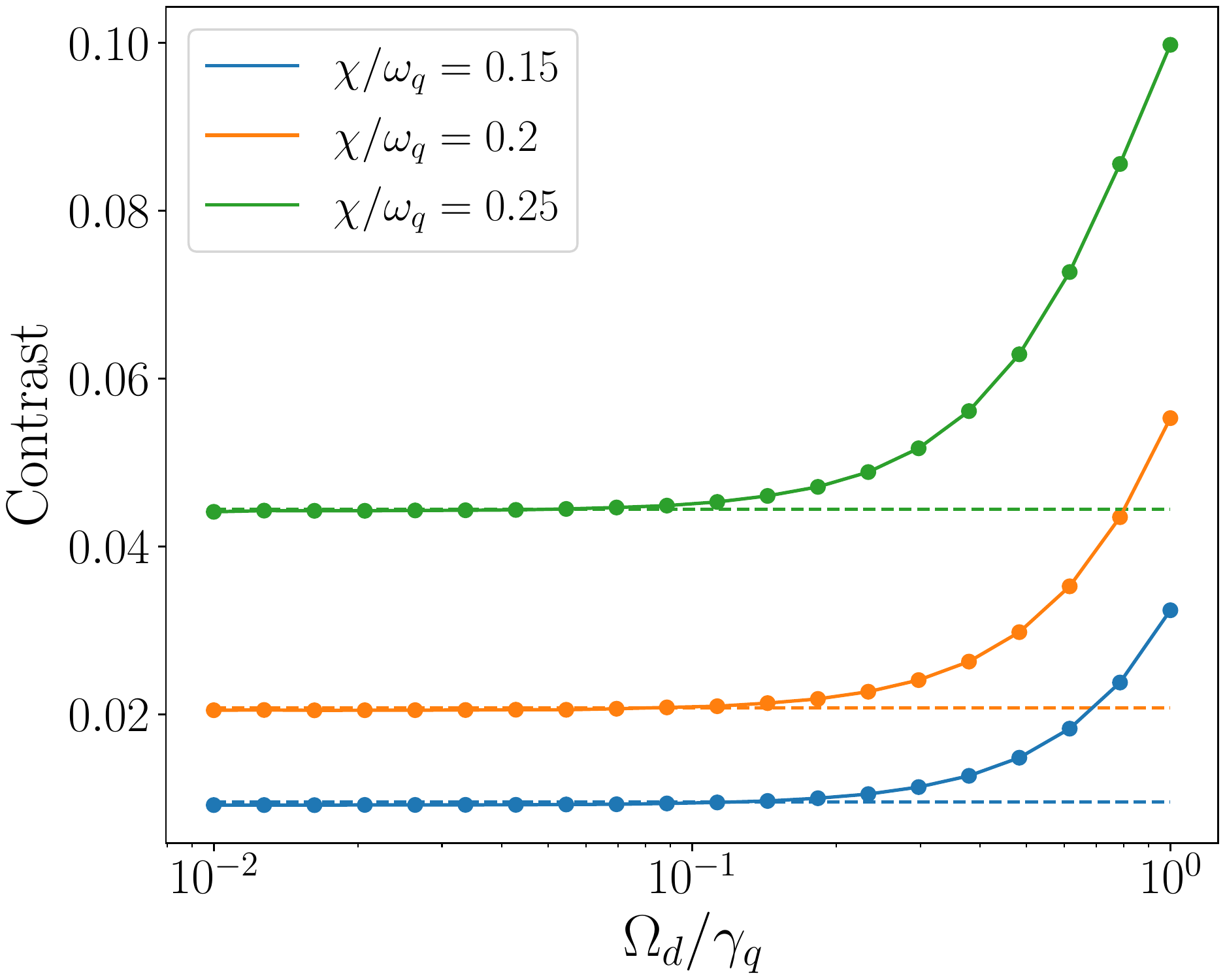} 
		\caption{The contrast between the \ak{peaks} at $\omega_2$ and $\omega_0$ is plotted against the Rabi frequency $\Omega_d$ for three different values \ak{of} the direct dispersive coupling strength $\chi/\omega_q = 0.15$, $\chi/\omega_q = 0.2$ and $\chi/\omega_q = 0.25$. The bold lines represent the contrast generated from simulations and the dashed lines the corresponding theory value for the contrast. The remaining parameters are fixed by $\omega_\alpha/\omega_q = 0.5$, $r=0.2$ and $\gamma_q/\omega_q = 0.1$. 
			\label{fig:Fig_supp1}}
	\end{figure}
	
	\begin{figure}[tb]
		\centering
		\includegraphics[width=1\columnwidth]{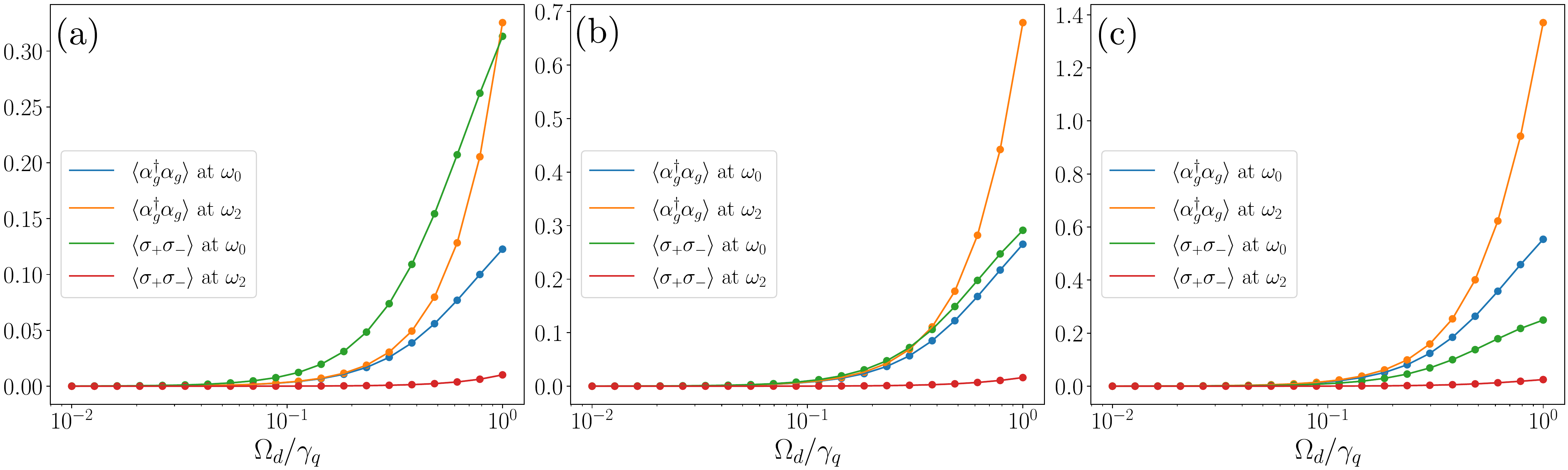} \caption{The steady state ground state squeezed-magnon number evaluated at the system resonance frequencies $\omega_0$ and $\omega_2$ as well as the qubit population, also evaluated at $\omega_0$ and $\omega_2$, are plotted against the Rabi frequency for three values of direct dispersive coupling strength (a) $\chi/\omega_q = 0.15$, (b) $\chi/\omega_q = 0.2$ and (c) $\chi/\omega_q = 0.25$.  The remaining parameters are fixed by $\omega_\alpha/\omega_q = 0.5$, $r=0.2$ and $\gamma_q/\omega_q = 0.1$. 
			\label{fig:Fig_supp2}}
	\end{figure}

	\subsection{Resolving Further Peaks $\omega_{2n}$ \label{subsec:further_peaks}}
	In the main text, we address how to resolve the first non-trivial peak at resonance frequency $\omega_2$. In the following, we want to address the experimental subtleties when resolving the further peaks at frequency $\omega_{2n}$ associated with the transition $\ket{0}_g \rightarrow \ket{2n}_e$ for $n>1$. Since the excitation probability $p_{2n}$ is suppressed with larger $n$, the amplitudes of the  corresponding excitation peaks in the qubit spectroscopy \ak{decrease by the evaluated contrast [Eq.~(5)] from one peak to the next}. \ak{The main point is that since these peaks are well-separated in frequency, we can use larger Rabi drives for probing subsequent peaks without violating the excitation linearity discussed above. We demonstrate this point by simulating the qubit spectroscopy of $\omega_4$ peak explicitly.} 
	
	\akcom{I have removed an entire paragraph since I found it to be unnecessary.}

	%The theory value for the contrast between the second non-trivial peak and the peak at $\omega_0$ is given by $c^\mathrm{(second~peak)}= |c_4|^2/|c_0|^2$ with the $c_{2n}$ from Eq~(5). This can be evaluated for small $\left|\chi\right|\ll\min\left[\left|A\right|,\left|A\left(A/2\left|B\right|-2\left|B\right|/A\right)\right|\right]$ as 
	%\begin{equation}
	%	c^\mathrm{(second~peak)} \approx 6\chi^4 \frac{|B|^4}{\left( A^2 -4|B|^2 \right)^4}, \label{eq:contrast_4}
	%\end{equation}
	%and for small $\left|B\right|\ll\min\left[\left|A-\chi\right|,\left|A+\chi\right|\right]$ the contrast can be expanded as $c^\mathrm{(second~peak)}\approx 6\chi^{4}\left|B\right|^{4}/\left(A^{2}-\chi^{2}\right)^{4}$. Since the excitation $\ket{0}_g \rightarrow \ket{2n}_e$ is suppressed by $p_4 = |c_4|^2$, the condition for the linear response regime $ \Omega_d^2 \ll \gamma_q^2/8 p_{4}$ ``allows'' for larger $\Omega_d$ enhancing the amplitude of the excitation peak at resonance frequency $\omega_4$. We therefore suggest to probe the non-trivial peaks at larger $\Omega_d$.

	In Fig.~\ref{fig:Fig_supp3} we show the steady state qubit excitation $\langle \sigma_+ \sigma_-\rangle_{\mathrm{st}}$ at resonance frequencies $\omega_0$, $\omega_2$ and $\omega_4$. In the simulations, we probe each of the excitation peaks \ak{employing} different Rabi frequencies $\Omega_d$: $\Omega_d = \gamma_q/7$ for $\omega_0$, $\Omega_d = 3\gamma_q/7$ for $\omega_2$ and $\Omega_d = 9\gamma_q/7$ for $\omega_4$. \ak{Since the qubit excitation scales as $\sim \Omega_d^2$ in the linear response regime [Eq.~\eqref{eq:qustead}]}, we can define the experimental contrast for the first non-trivial peak
	\begin{equation}
		c_\mathrm{experiment} = \frac{\left\langle \hat{\sigma}_{+}\hat{\sigma}_{-}\right\rangle _{\text{st}}\left( \omega_d=\omega_2 \right)/\Omega_{d,2}^2}{\left\langle \hat{\sigma}_{+}\hat{\sigma}_{-}\right\rangle _{\text{st}}\left( \omega_d=\omega_0 \right)/\Omega_{d,0}^2},
	\end{equation}
	where $\Omega_{d,2n}$ denotes the Rabi frequency used to resolve the peak at $\omega_{2n}$ for $n=0,1$. An analogous expression can be defined for the contrast of the second non-trivial peak. \ak{The simulation reported in Fig.~\ref{fig:Fig_supp3} shows that employing a larger Rabi drive for the non-overlapping peaks allows to correctly obtain the contrast due to the equilibrium superposition from even the weak peaks.} As a last note, the resolution of the peaks becomes better with smaller qubit dissipation $\gamma_q$ since the overlap between excitation peaks decreases.
	
	\begin{figure}[tb]
		\centering
		\includegraphics[width=1\columnwidth]{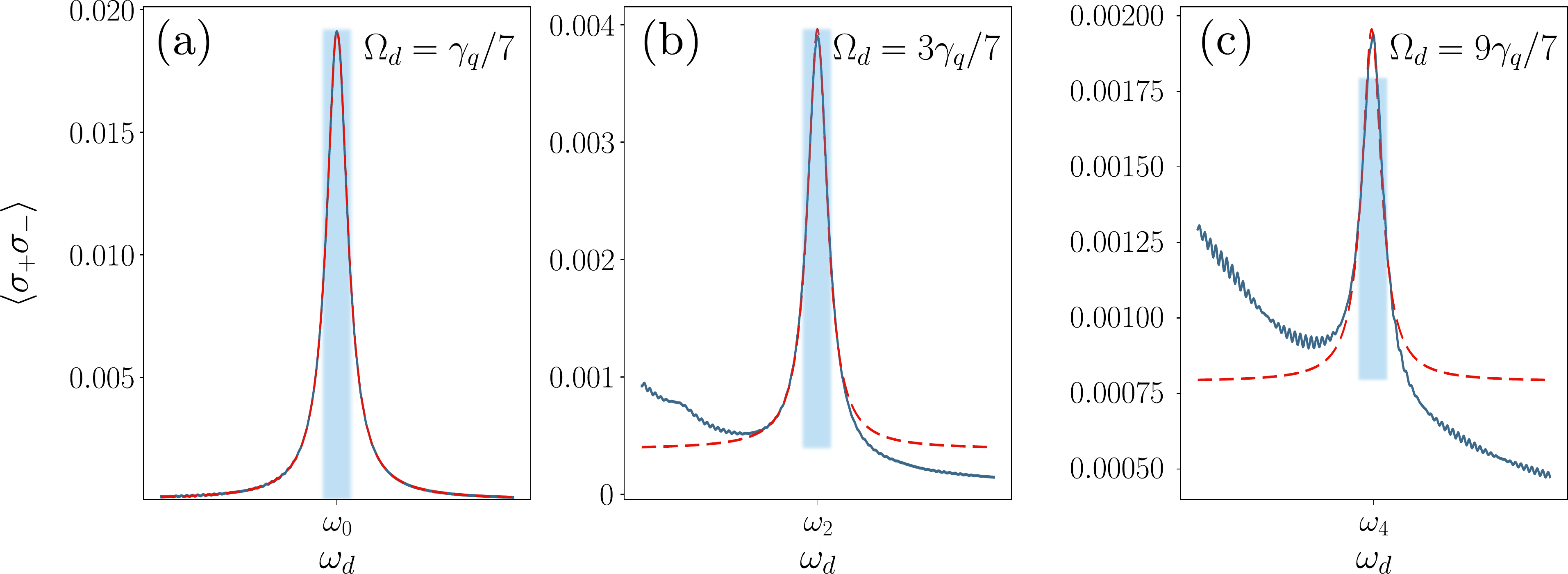} \caption{Steady state qubit population vs. drive frequency. The \ak{spectra} around the peaks at $\omega_0$, $\omega_2$ and $\omega_4$ \ak{are} simulated with different values of the Rabi frequency: $\Omega_d = \gamma_q/7$ for $\omega_0$, $\Omega_d = 3\gamma_q/7$ for $\omega_2$ and $\Omega_d = 9\gamma_q/7$ for $\omega_4$. The bold lines are the simulation data, the dashed lines the Lorentzian fits and the bars represent the theory values. The theory bars have been multiplied by $\Omega_d^2$ for $\omega_2$ and $\omega_4$ \ak{to account for the larger Rabi drives employed}. 
			\label{fig:Fig_supp3}}
	\end{figure}
	
}

\section{Coherent Coupling in the Detuned Limit \label{sec:coherent_coupling}}
%Introduction and goals
Spin qubits couple dispersively and coherently with the magnet (see
section \ref{subsec:H_dis}). Superconducting qubits on the other hand
couple coherently to the magnonic mode and have no ``in-built'' direct
dispersive coupling to magnets~\citep{tabuchi2016,tabuchi2015}.
However, one obtains an effective dispersive interaction arising from the coherent
coupling in the highly detuned limit~\cite{schuster2007,boissonneault2009}. This is why we want to discuss the
coherent coupling in the dispersive limit and address the question
if the magnon composition of an equilibrium squeezed magnetic ground
state can be resolved with this kind of interaction as well. We also
want to discuss if the measurement mechanism
obtained from direct dispersive coupling is perturbed when coherent coupling is present. This effective dispersive interaction has been detailed in the text book~\citep{gerry2005} or, for instance,
in~\citep{boissonneault2009} and \citep{schuster2007}.

%Description: Hamiltonian and effective frequency shift
Let's consider a ferromagnet with anisotropies as discussed section \ref{subsec:H_FM}. We now assume that the local magnon $\hat{a}$
(``spin flip'') and a qubit $\hat{\sigma}$ are coupled coherently
via Rabi interaction~\citep{rabi1936,rabi1937} and that there is
no direct dispersive coupling. The corresponding system Hamiltonian
has the form 
\begin{align}
	\hat{\mathcal{H}}_{\text{sys, SC}}=A\hat{a}^{\dagger}\hat{a}+B\hat{a}^{2}+B^{*}\hat{a}^{\dagger2}+\frac{\omega_{q}}{2}\hat{\sigma}_{z}+g\left(\hat{a}^{\dagger}+\hat{a}\right)\left(\hat{\sigma}_{+}+\hat{\sigma}_{-}\right).\label{eq:H_sc_supp}
\end{align}
where the coherent coupling strength is denoted by $g$. We transform
into the eigenbasis of the ferromagnet by using the Bogoliubov transformation
{[}Eq.~\eqref{eq:Bogoliubov}{]} such that the system Hamiltonian
becomes
\begin{align}\label{eq:HSCalpha}
	\hat{\mathcal{H}}_{\text{SC,\ensuremath{\alpha}}}=\omega_{\alpha}\hat{\alpha}^{\dagger}\hat{\alpha}+\frac{\omega_{q}}{2}\hat{\sigma}_{z}+\tilde{g}\hat{\alpha}^{\dagger}\hat{\sigma}_{-}+\tilde{g}^{*}\hat{\alpha}\hat{\sigma}_{+}+\tilde{g}\hat{\alpha}^{\dagger}\hat{\sigma}_{+}+\tilde{g}^{*}\hat{\alpha}\hat{\sigma}_{-},
\end{align}
where the modified coupling strength reads $\tilde{g}=g\left(\cosh r-e^{i\theta}\sinh r\right)$
and we neglected the vacuum energy. The dispersive limit is defined
by a large detuning of the boson and the qubit while being relatively
weakly coupled such that $\tilde{g}\ll\left|\omega_{q}-\omega_{\alpha}\right|$.
This way the modes do not hybridize and the interaction can be treated
perturbatively. Now if we do not neglect the fast rotating terms (taking
into account the full Hamiltonian $\hat{\mathcal{H}}_{\text{SC},\alpha}$
{[}Eq.~\eqref{eq:HSCalpha}{]}) performing perturbation theory
will lead to a diagonal term $\propto\hat{\alpha}^{\dagger}\hat{\alpha}$
and non-diagonal terms $\propto\hat{\alpha}^{2},\left(\hat{\alpha}^{\dagger}\right)^{2}$.
Assuming $\tilde{g}$ to be real, \citet{zueco2009} have shown that
the dispersive limit beyond the rotating wave approximation leads
to an effective frequency shift in the qubit frequency 
\begin{align}
	\tilde{\chi} & =\tilde{g}^{2}\left(\frac{1}{\omega_{q}-\omega_{\alpha}}+\frac{1}{\omega_{q}+\omega_{\alpha}}\right). \label{eq:effective_chi_supp}
\end{align}

\begin{figure}[tb]
	\centering\includegraphics[width=0.6\columnwidth]{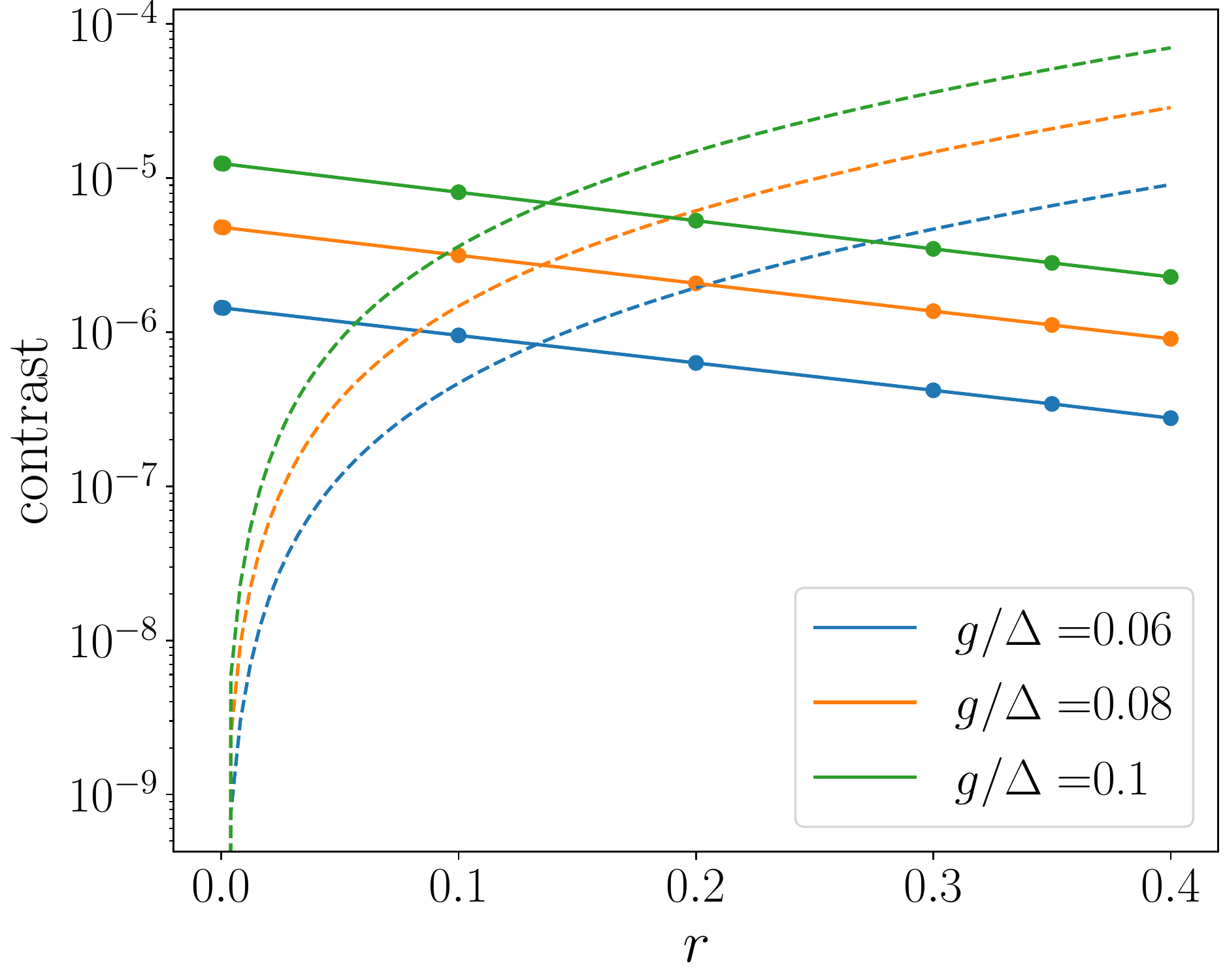} \caption{The contrast between the first non-trivial peak and the trivial peak
		is plotted as a function of the squeezing $r$ for several values
		of coherent coupling $g$. We compare simulation data (points and
		solid lines) with the expected contrast arising from the analytic
		model with direct dispersive coupling $\chi = \tilde{\chi}$ [Eq.~\eqref{eq:effective_chi_supp}]. The fixed parameters are $\omega_{\alpha}/2\pi=5\,\mathrm{GHz}$, $\omega_{q}/2\pi=10\,\mathrm{GHz}$,
		$\gamma_{q}/2\pi=10\,\mathrm{MHz}$ and $\Omega_{d}/2\pi=1.4\,\mathrm{MHz}$.
		\label{fig:Fig_supp4}}
\end{figure}

%Simulations with chi=0
The Hamiltonian $\hat{\mathcal{H}}_{\text{sys,SC}}$ {[}Eq.~(\ref{eq:H_sc_supp}){]}
cannot be easily treated analytically. For this reason, we perform
the numerical simulations of the qubit spectroscopy as described in
the main text and the SM. We keep the bare squeezed-magnon frequency
$\omega_{\alpha}/2\pi=5\,\mathrm{GHz}$ and the qubit frequency $\omega_{q}/2\pi=10\,\mathrm{GHz}$.
For higher precision, we choose a small qubit decay rate $\gamma_{q}/2\pi=10\,\mathrm{MHz}$
and Rabi frequency $\Omega_{d}/2\pi=1.4\,\mathrm{MHz}.$ We find that
there is a non-trivial peak arising around $\omega_{d}\approx\omega_{q}+2\omega_{\alpha}$
which we analyse varying the squeezing $r$ and the coherent coupling
$g$. In Fig.~\ref{fig:Fig_supp4}, we show the contrast plotted
against the squeezing $r$ for several values of $g$. We compare
the contrast resulting from the numerical simulations (points and
solid lines) with the contrast arising from the analytic theory with
direct dispersive coupling (dashed lines) given by Eq.~(5) from the
main text. We find that the numerical calculation with coherent coupling and the analytic
model arising from direct dispersive coupling $\chi = \tilde{\chi}$ [Eq.~\eqref{eq:effective_chi_supp}] do not match. While
the contrast in our analytic model is equal to zero in the absence of squeezing and increasing with larger squeezing $r$, the contrast
in the numerical data is nonzero for $r=0$ and decreasing with larger
$r$. We therefore conclude that the peak is unrelated to the measurement
mechanism obtained with direct dispersive coupling and stems from
another higher order process arising from the counter rotating terms.
This shows that the direct dispersive coupling $\chi$ is essential for resolving
the nonclassical magnon number composition of the magnetic ground
state. The previous method~\cite{schuster2007} for resolving nonequilibrium superpositions using an effective dispersive coupling $\tilde{\chi}$ does not work for equilibrium nonclassical superpositions~\cite{kamra2020}.

\begin{figure}
	\centering \includegraphics[width=0.6\columnwidth]{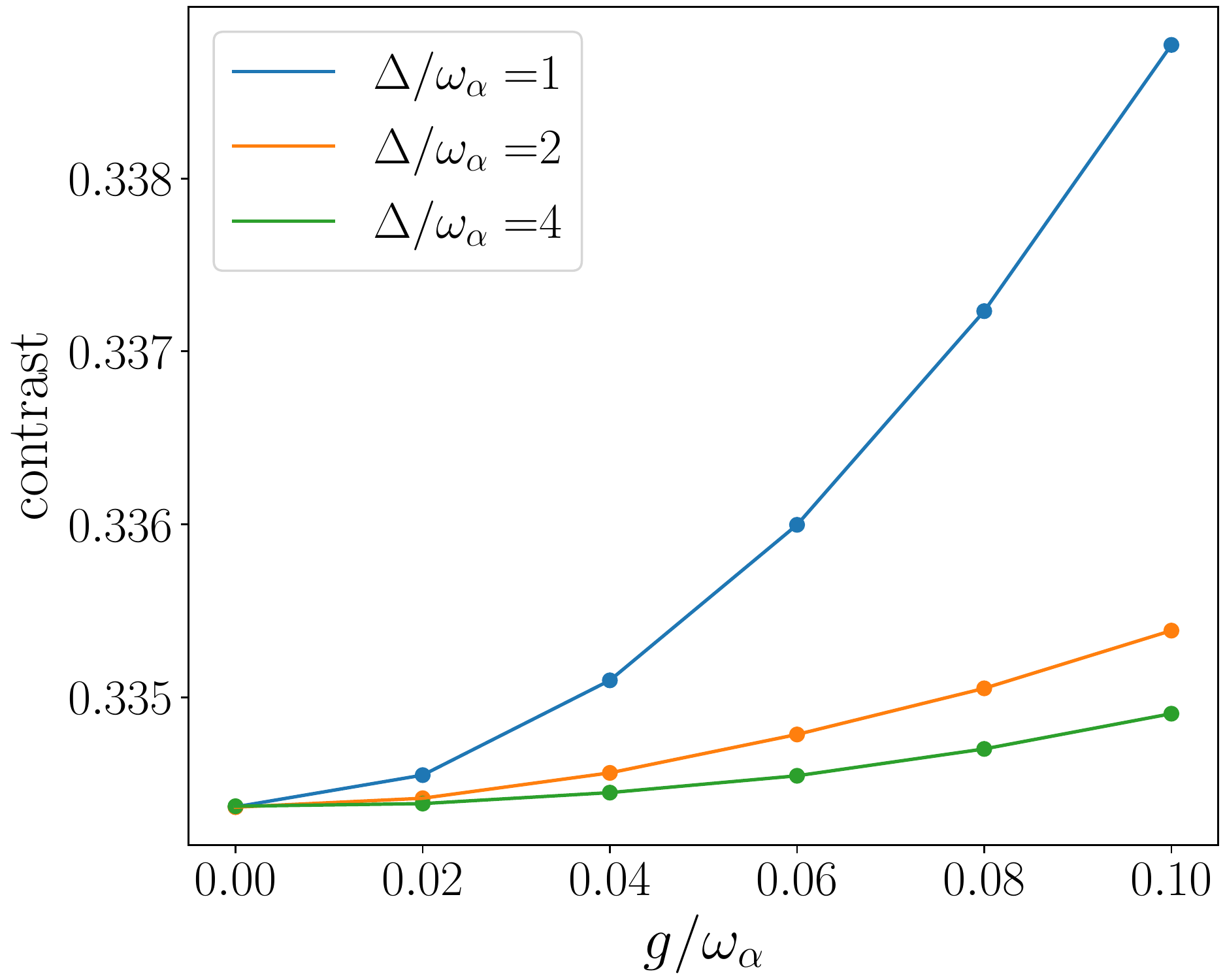} \caption{The contrast between the first non-trivial peak and the trivial peak
		is plotted as a function of coherent coupling $g$ for several values
		of magnon-qubit detuning $\Delta=\omega_{q}-\omega_{\alpha}$. The
		data are obtained by simulating the qubit spectra under the evolution
		of $\hat{\mathcal{H}}_{\text{full}}$ [Eq.~\eqref{eq:H_dis_coh}] and determining the height of
		the two peaks via Lorentzian fits. We fix the simulation parameters
		by $\omega_{\alpha}/2\pi=5\,\mathrm{GHz}$, $r=0.45$, $\chi/2\pi=2\,\mathrm{GHz}$,
		$\gamma_{q}/2\pi=1\,\mathrm{GHz}$ and $\Omega_{d}/2\pi=0.14\,\mathrm{GHz}$.
		\label{fig:Fig_supp5}}
\end{figure}

%Simulations with chi=2
Lastly, we want to explore what happens when a qubit is coupled via direct
dispersive interaction and coherent interaction. Thereby, we address
the question if the coherent coupling perturbs the measurement mechanism
obtained from direct dispersive coupling when the bare squeezed-magnon
frequency $\omega_{\alpha}$ and the bare qubit frequency $\omega_{q}$
are far detuned. For this reason, we consider the full Hamiltonian

\begin{align}
	\hat{\mathcal{H}}_{\text{full}}=A\hat{a}^{\dagger}\hat{a}+B\hat{a}^{2}+B^{*}\hat{a}^{\dagger2}+\frac{\omega_{q}}{2}\hat{\sigma}_{z}+\chi\hat{a}^{\dagger}\hat{a}\hat{\sigma}_{z}+g\left(\hat{a}^{\dagger}+\hat{a}\right)\left(\hat{\sigma}_{+}+\hat{\sigma}_{-}\right),\label{eq:H_dis_coh}
\end{align}
which contains the ferromagnet with anisotropies, the qubit, and both direct and coherent coupling between the magnon $\hat{a}$ and the qubit.
Since the Hamiltonian $\hat{\mathcal{H}}_{\text{full}}$ is not analytically
tractable, we analyse it numerically using our established qubit spectroscopy
simulations as discussed in section \ref{sec:simulation_supp}. In Fig.~\ref{fig:Fig_supp5},
we show the contrast between the first non-trivial peak and the trivial peak arising from
the numerical simulations as a function of $g$ at different values of the magnon-qubit
detuning $\Delta=\omega_{q}-\omega_{\alpha}$. We vary the detuning by changing the bare qubit frequency $\omega_q$. The rest of the parameters
are fixed by $\omega_{\alpha}/2\pi=5\,\mathrm{GHz}$, $r=0.45$, $\chi/2\pi=2\,\mathrm{GHz}$,
$\gamma_{q}/2\pi=1\,\mathrm{GHz}$ and $\Omega_{d}/2\pi=0.14\,\mathrm{GHz}$.
For a detuning of $\Delta=5\,\mathrm{GHz}$ the deviation in the contrast caused by
the coherent coupling is less than $1.5\,\%$ for $g=0.5\,\mathrm{GHz}$
which is already at the limit between weak coupling and strong coupling regime. For a larger detuning
of $\Delta=10\,\mathrm{GHz}$ and $\Delta=20\,\mathrm{GHz}$ the curves are more flat in the range of $g$ that we consider. We therefore conclude that the
perturbation caused by the coherent coupling is insignificant for weak coupling
strengths $g$ and that the influence of the coherent coupling can be suppressed using larger detuning $\Delta$.

\al{
	\ak{\section{Experimental platforms for direct dispersive coupling }}
	
	In \ak{this} section, we \ak{examine some available experimental setups}
	that \ak{may achieve the} direct dispersive coupling between a qubit and magnetic insulator. In section \ref{subsec:H_dis} \ak{above}, we \ak{have derived} the
	direct dispersive interaction $\hat{\mathcal{H}}_{\text{dis}}=\chi\hat{a}^{\dagger}\hat{a}\hat{\sigma}_{z}$~[Eq.~(\ref{eq:hdisfinal})] \ak{starting with an exchange coupling Hamiltonian}.
	The \ak{direct dispersive term} arises from the $z$-components of the spin-spin interaction
	$\propto\hat{S}_{z}\hat{\sigma}_{z}$ and \ak{its strength is given by}
	\begin{align}
		\chi & =\frac{J_{\text{int}}N_{\text{int}}}{2N_{\text{F}}}\left|\psi\right|^{2},
	\end{align}
	with the interfacial \ak{exchange} coupling strength $J_{\text{int}}$, the number of
	interfacial lattice sites $N_{\text{int}}$, the number of lattice sites
	in the ferromagnet $N_{\text{F}}$ and the averaged spin qubit wave function
	$\left|\psi\right|^{2}$. 
	
	\ak{Here, we consider} two types of qubits that provide the interaction $\propto\hat{S}_{z}\hat{\sigma}_{z}$ \ak{noting that better candidates might have escaped our attention or become available in the near future}. First, we focus on semiconducting quantum dots that interact through exchange coupling
	with magnets~\cite{chatterjee2021}. We then turn our attention to nitrogen-vacancy (NV)
	defects in diamond \ak{which may} interact with magnets via magnetic dipole-dipole coupling~\cite{casola2018}. From experimentally known parameters and set-ups, we \ak{provide design equations for optimizing these platforms and achieving a desired} $\chi$.
	
	\subsection{Exchange interaction}
	
	Semiconducting quantum
	dots implement spin exchange interaction\ak{, see Ref.~\cite{chatterjee2021} for a review. Such a magnet--spin qubit system focusing on coherent interaction has been detailed in Ref.~\cite{skogvoll2021}. Considering }this set-up, \ak{we} evaluate the direct dispersive interaction between far detuned spin qubit and magnon. Assuming the qubit wave function to be localized in $N_{\text{layer}}$ monolayers of an equally thin ferromagnet with $N_{\text{int}}$ interfacial sites, such that $\left|\psi\right|^{2} = 1/N_\mathrm{F}$ with $N_\mathrm{F} = N_\mathrm{layers} N_\mathrm{int}$, the direct dispersive coupling strength becomes \akcom{This discussion is not self-contained. It has not been mentioned how one obtains the following equation by making assumptions about the qubit wavefunction. In the interest of not being slowed down by such details, I suggest that we accept it and move on. An interested reader will need to read Ida's paper anyway. However, please consider this comment as a lesson for future writing.}
	\begin{align}\label{eq:chies}
		\chi & = \frac{J_{\text{int}}}{2N_{\text{layers}}^{2}N_{\text{int}}},
	\end{align}
	which is increasing with decreasing size of the magnet. From spin-pumping experiments~\cite{kajiwara2010, czeschka2011, weiler2013}, we take the interfacial exchange coupling $J_{\text{int}}\approx10\,\mathrm{meV}$
	for our estimation.  Assuming $N_{\text{layer}}=5$ monolayers and
	$N_{\text{int}}=1000$ interfacial sites yields $\chi=0.00002\cdot J_{\text{int}}=0.3\,\text{GHz}$. Reducing the number of interfacial sites to $N_{\text{int}}=100$
	results in an increased direct dispersive coupling strength $\chi=3\,\mathrm{GHz}$. \ak{In conclusion, Eq.~\eqref{eq:chies} above provides the necessary design equation that could be employed in engineering a desired value of the direct dispersive coupling. It also shows that smaller system sizes on the nanoscale are needed for this.}
	
	\subsection{Dipole-dipole coupling}
	
	\akcom{What is the purpose of mentioning the NV center spin? I think there is none. On the other hand, an uninformed reader might wonder why an NV center is a qubit when it has three states corresponding to spin of 1.} NV center \ak{defects in diamond interact} with other spins through
	dipole-dipole interaction~\cite{casola2018}. In the supplemental \ak{information of Ref.}~\cite{schlipf2017}, the interaction between NV center and network spins is modeled by a term $\propto\hat{S}_{z}\hat{\sigma}_{z}$. This motivates us to examine the suitability of NV centers as spin qubits for our measurement protocol.
	
	A general expression of magnetic dipole-dipole interaction between an NV center and a magnetic insulator reads
	\begin{align}
		\hat{\mathcal{H}}_{\text{dd}} & =\frac{\mu_{0}}{4\pi}\sum_{l}\frac{1}{r_{l}^{3}}\left[\boldsymbol{M}_{l}\cdot\boldsymbol{m}_{\text{NV}}-3\left(\boldsymbol{M}_{l}\cdot\hat{\boldsymbol{r}}_{l}\right)\left(\boldsymbol{m}_{\text{NV}}\cdot\hat{\boldsymbol{r}}_{l}\right)\right],\label{eq:H_dd}
	\end{align}
	where $\boldsymbol{m}_{\text{NV}}$ is the magnetic moment of the NV center, $\boldsymbol{M}_{l}$ denotes the magnetic moment of interfacial lattice site $l$ and $r_{l}$ is the distance between the NV center and interfacial lattice site $l$.  $\hat{\boldsymbol{r}}_{l}$ is a unit vector in the direction of the line joining $\boldsymbol{m}_{\text{NV}}$ and $\boldsymbol{M}_{l}$. We want to bring Eq.~(\ref{eq:H_dd}) into the form of Eq.~(\ref{eq:H_int}) in order to estimate
	an effective \ak{$J_{\mathrm{int}}$} from dipole-dipole interaction which we denote by $J_{\text{dd}}$. \ak{This is accomplished by positioning the NV center such that its spin and displacement from the magnet are orthogonal to each other.} Assuming the magnetic moments $\boldsymbol{m}_{\text{NV}}$ and $\boldsymbol{M}_{l}$ to be in plane and the unit
	vectors $\hat{\boldsymbol{r}}_{l}$ to be out of plane, Eq.~(\ref{eq:H_dd}) becomes 
	\begin{align}
		\hat{\mathcal{H}}_{\text{dd}} & =\frac{\mu_{0}}{4\pi}\sum_{l}\frac{\gamma_{l}\gamma_{\text{NV}}}{r_{l}^{3}}\hat{\boldsymbol{S}}_{l}\cdot\boldsymbol{\hat{s}}_{\text{NV}}, \label{eq:H_dd_red}
	\end{align}
	where $\hat{\boldsymbol{S}}_{l}$ and $\boldsymbol{\hat{s}}_{\text{NV}}$
	denote the spin operators at lattice site $l$ and the NV center, and $\gamma_{l}$ and $\gamma_{\text{NV}}$ are the corresponding gyromagnetic ratios. Finally, assuming that the
	distances $r_l$ are approximately equal $r_{l}\approx r$ and that the gyromagnetic ratios of the interfacial sites are the same $\gamma_{l}=\gamma$, we obtain the effective coupling
	\begin{align}
		J_{\text{dd}} & =\frac{\mu_{0}}{4\pi}\frac{\gamma\gamma_{\text{NV}}}{r^{3}},
	\end{align}
	which strongly depends on the distance $r$. The distance between
	the probe and the NV center in experiments can be around $r=100\,\text{nm}$~\cite{lee-wong2020}
	down to tens of nanomometers~\cite{maze2008,vandersar2015}. For electron spins $\gamma_{l}=\gamma_{\text{NV}}=\gamma_{e}$,
	we estimate an effective interfacial coupling strength of $J_{\text{dd}}\approx0.16\,\text{peV}$
	for a distance of $r=100\,\text{nm}$ and $J_{\text{dd}}\approx0.16\,\text{neV}$
	for $r=10\,\text{nm}$. Even the larger value $J_{\text{dd}}\approx0.16\,\text{neV}$
	is $10$ orders smaller than the interfacial coupling $J_{\text{int}}$
	from exchange interaction. This would result in a direct dispersive
	coupling strength in the order of $\chi\propto0.1\,\text{Hz}$. We
	conclude that this coupling strength is too weak and magnetic dipole-dipole interaction is not suitable for our purposes.
}

\ak{However, newly developed protocols and qubits employing electron spin resonance (ESR) and single spins offer the potential for smaller distances. In such studies, the magnetic dipolar interactions have been found to play an important role and may admit significant direct dispersive interactions~\cite{yang2019,veldman2021,choi2017}.} \akcom{Please cite the additional relevant paper you found on such ESR studies.}

\end{document}